\documentclass[preprint,aps,floatfix,nofootinbib,a4paper, superscriptaddress]{revtex4-2}
\pdfoutput=1
\usepackage{amsmath, amssymb, amsfonts, amsthm, latexsym, epsfig, mathrsfs, xcolor, bbm, slashed}

\usepackage[inline]{enumitem}

\usepackage{setspace}
\usepackage[marginal, multiple]{footmisc}
\usepackage[T1]{fontenc}
\usepackage[utf8]{inputenc}
\usepackage{lmodern}
\usepackage{mathtools}
\usepackage[colorlinks, allcolors=blue!70!black, linktocpage]{hyperref}
\numberwithin{equation}{section}

\usepackage{cleveref}

\usepackage{microtype}

\usepackage{floatrow}
\let\OLDtableofcontents\tableofcontents
\renewcommand\tableofcontents[1]{%
    {\baselineskip 0.5ex %
	\OLDtableofcontents{#1}}%
}

\let\OLDthebibliography\thebibliography
\renewcommand\thebibliography[1]{%
	\setstretch{1.079} 
	\OLDthebibliography{#1}%
	\small %
	\setlength{\itemsep}{0.2\baselineskip} 
}

\let\OLDfootnote\footnote
\renewcommand\footnote[1]{%
	\setlength{\footnotesep}{0.75\baselineskip}%
	{\footnotesize \OLDfootnote{#1}}%
}

\setlist[enumerate]{noitemsep, label=(\arabic*), ref=(\arabic*)}

\newlist{condlist}{enumerate}{2}
\setlist[condlist,1]{noitemsep, topsep=0pt, label=(\arabic*), ref=(\arabic*)}
\setlist[condlist,2]{noitemsep, label=(\alph*), ref=(\arabic{condlisti}.\alph*)}
\crefname{condlisti}{condition}{conditions}
\crefname{condlistii}{condition}{conditions}

\newlist{propertylist}{enumerate}{1}
\setlist[propertylist,1]{noitemsep, topsep=0pt, label=(\arabic*), ref=(\arabic*)}
\crefname{propertylisti}{Property}{Properties}

\renewcommand\thesection{\arabic{section}}
\renewcommand\thesubsection{\arabic{subsection}}

\makeatletter
\def\p@subsection{\thesection.}
\def\p@subsubsection{\thesection.\thesubsection.}
\makeatother 

\theoremstyle{plain}

\theoremstyle{definition}
\newtheorem{definition}{Definition}[section]

\theoremstyle{remark}
\newtheorem{remark}{Remark}[section]

\crefname{equation}{Eq.}{Eqs.}
\creflabelformat{equation}{#2#1#3}

\crefname{section}{\S}{\S}
\crefname{appendix}{Appendix}{Appendices}
\crefname{figure}{Fig.}{Figs.}
\crefname{table}{Table}{Tables}
\crefname{paragraph*}{Para.}{Paras.}

\crefname{definition}{Def.}{Defs.}
\crefname{prop}{Prop.}{Props.}
\crefname{lemma}{Lemma}{Lemmas}
\crefname{corollary}{Cor.}{Cors.}
\crefname{thm}{Theorem}{Theorems}
\crefname{remark}{Remark}{Remarks}

\crefname{ass}{Assumptions}{Assumptions}
\crefname{property}{Properties}{Properties}

\newcommand{\be}{\begin{equation}\begin{aligned}}
\newcommand{\ee}{\end{aligned}\end{equation}}

\newcommand{\lb}{\left}
\newcommand{\rb}{\right}
\newcommand{\scrp}{\scri^+}

\newcommand{\mc}{\mathcal}

\newcommand{\ms}{\mathscr}
\newcommand{\mf}{\mathfrak}
\newcommand{\bb}{\mathbb}

\DeclareMathOperator{\STF}{STF}

\newcommand{\eqsp}{\, ,\quad} 

\newcommand{\hr}{\begin{center}* * *\end{center}}

\newcommand{\Lie}{\pounds} 
\newcommand{\defn}{\mathrel{\mathop:}=} 

\newcommand{\union}{\cup} 
\newcommand{\inter}{\cap} 

\let\oldsetminus\setminus
\renewcommand{\setminus}{\!\oldsetminus\!} 

\let\oldint\int
\renewcommand{\int}{\oldint\limits}

\let\oldlim\lim
\renewcommand{\lim}{\oldlim\limits}

\renewcommand{\bar}{\overline}

\newcommand{\scri}{\ms I}

\newcommand{\hyp}{\ms H}

\newcommand{\cyl}{\ms C}
\newcommand{\nulls}{\ms N}

\newcommand{\hateq}{\mathrel{\mathop {\widehat=} }} 

\newcommand{\dd}[1]{\boldsymbol{#1}} 

\newcommand{\df}[1]{#1} 

\newcommand{\nfrac}[2]{{{}^#1\!\!/\!_#2}}

\newcommand{\half}{\nfrac{1}{2}}

\newcommand{\Omh}{\Omega^\half} 

\let\eth\relax
\DeclareMathOperator{\eth}{\text{\rm \dh}}

\newcommand{\wt}{\circeq}

\newcommand{\dv}[1]{{}^\star\!{\dd{#1}}}
\begin{document}

\setstretch{1.2}


\title{Conservation of asymptotic charges from past to future null infinity: Lorentz charges in general relativity}

\author{Kartik Prabhu}
\email{kartikprabhu@ucsb.edu}
\affiliation{Department of Physics, University of California, Santa Barbara, CA 93106, USA}
\author{Ibrahim Shehzad}
\email{is354@cornell.edu}
\affiliation{Department of Physics, Cornell University, Ithaca, NY 14853, USA}

\begin{abstract}
We show that the asymptotic charges associated with Lorentz symmetries on past and future null infinity match in the limit to spatial infinity in a class of asymptotically-flat spacetimes. These are spacetimes that obey the Ashtekar-Hansen definition of asymptotic flatness at null and spatial infinity and satisfy an additional set of conditions which we lay out explicitly. Combined with earlier results on the matching of supertranslation charges, this shows that \emph{all} Bondi-Metzner-Sachs (BMS) charges on past and future null infinity match in the limit to spatial infinity in this class of spacetimes, proving a relationship that was conjectured by Strominger. Assuming additional suitable conditions are satisfied at timelike
infinities, this proves that the flux of all BMS charges is conserved in any classical
gravitational scattering process in these spacetimes.
\end{abstract}

\maketitle
\newpage
\tableofcontents

\section{Introduction}
\label{sec:intro}

In general relativity in four dimensions, the asymptotic symmetry groups at past and future null infinity in asymptotically-flat spacetimes are the (a priori independent) infinite-dimensional Bondi-Metzner-Sachs (BMS) groups (often denoted by BMS$^-$ and BMS$^+$ respectively) \cite{BBM, Sachs1} (see also \cite{Ashtekar:2014zsa, GPS} for recent reviews). It is natural to ask how these groups are related in the limit to spatial infinity along past and future null infinity. It has in fact been conjectured by Strominger \cite{Stro-CK-match} that the generators of these groups match (up to antipodal reflection) in the limit to spatial infinity and, moreover, that the associated charges on cross-sections of past and future null infinity are equal in this limit. If this matching of symmetries and charges can be proven, it would imply the existence of a \emph{global} ``diagonal'' asymptotic symmetry group for classical general relativity and the existence of an infinite number of conservation laws in classical gravitational scattering. The content of these conservation laws would be that for each generator of BMS$^-$ and its corresponding generator\footnote{Here, by ``corresponding'' we mean the generator of BMS$^+$ which this generator of BMS$^-$ matches up to antipodal reflection in the limit to spatial infinity where this limit is dictated by the equations that govern BMS symmetries on $\scri$ (\cref{eq:symm-null}).} in BMS$^+$, the difference of the associated charges evaluated on cross-sections of past and future null infinity would equal the difference of the incoming flux at past null infinity and the outgoing flux at future null infinity in the region between the two cross-sections (see \cref{eq:cons-law}). Moreover, if appropriate conditions are obeyed at timelike infinities such that the BMS charges all go to zero in the limit to timelike infinities (see Remark 4.6 of \cite{KP-GR-match}), for each such pair of identified generators of BMS$^-$ and BMS$^+$, the total incoming flux through past null infinity would equal the total outgoing flux through future null infinity.  Strominger has further conjectured that the diagonal asymptotic symmetry group obtained from the aforementioned matching of symmetries is the symmetry group of the scattering matrix in quantum gravity. In a series of related developments, the existence of these conservation laws has also been shown to be related to soft graviton theorems and gravitational memory effects, and has been conjectured to have implications for the black hole information paradox (see \cite{Strominger:2017zoo} and references therein; see also \cite{BP,Mirbabayi:2016axw} where the relevance of these conservation laws for the black hole information paradox was disputed).

The analogous matching of asymptotic symmetries and charges in Maxwell theory on asymptotically-flat spacetimes was shown in \cite{CE,KP-EM-match}. In the gravitational case, the BMS group is the semi-direct product of an infinite dimensional group of supertranslations with the Lorentz group. For asymptotic translations (which are an invariant subgroup of supertranslations) it was shown in \cite{Ash-Mag-Ash} that the Bondi \(4\)-momentum on past and future null infinity is equal to the ADM \(4\)-momentum in the limit to spatial infinity. Moreover, the matching of the all the asymptotic supertranslations and their charges was proven in linearized gravity around a Minkowski background spacetime in \cite{Tro,Mohamed:2021rfg}. The extension of these results to the matching of supertranslation charges in full nonlinear general relativity was shown in \cite{KP-GR-match}, assuming certain ``null regularity'' conditions on the behaviour of the relevant Weyl tensor components near spatial infinity. These ``null regularity'' conditions are satisfied in linearized gravity (as follows from \cite{Tro,Mohamed:2021rfg}) and the issue of their validity in general was discussed in \cite{KP-GR-match}.

The goal of this paper is to supplement the result of \cite{KP-GR-match} by proving the matching of Lorentz symmetries and their associated charges in asymptotically-flat spacetimes that satisfy certain assumptions which we specify in \cref{sec:AH}. For stationary spacetimes the matching of the Bondi and ADM angular momenta was shown in \cite{AS-ang-mom}. It was also argued in \cite{AS-ang-mom} that a similar analysis may hold in non-stationary spacetimes where the gravitational radiation falls-off towards spatial infinity. We will show that this is indeed true and specify the sufficient conditions that are needed. The main tool we will use in our analysis is the covariant formulation of asymptotic-flatness due to Ashtekar and Hansen \cite{AH} which treats both null and spatial infinities in a unified spacetime-covariant manner. In the Ashtekar-Hansen formalism, instead of working directly at spatial infinity, which is represented by a point, denoted by $i^0$, in the conformal-completion of any asymptotically-flat spacetime and where sufficiently smooth structure is unavailable (except in the case of Minkowski spacetime), one works with a ``blowup'' --- the space of spatial directions at \(i^0\) --- given by a timelike-unit-hyperboloid \(\hyp\) in the tangent space of \(i^0\). Suitably conformally rescaled fields whose limits to \(i^0\) depend on the direction of approach induce smooth fields on \(\hyp\) which can then be studied using standard differential calculus on \(\hyp\). Since we will be interested in considering limits of quantities defined on null infinity to spatial infinity, we will then conformally-complete this hyperboloid into a cylinder, $\ms{C}$, (as discussed in \cite{KP-EM-match,KP-GR-match}) whose boundaries \(\nulls^\pm\) are diffeomorphic to the space of (rescaled) null directions at spatial infinity. We then fix the supertranslation freedom in a neighborhood of $i^{0}$ to isolate a Lorentz subgroup of the BMS group. The generators of this Lorentz subgroup are Killing vector fields on \(\hyp\) and conformal Killing vector fields on the space of null directions \(\nulls^\pm\) at \(i^0\). The Killing equation on \(\hyp\) leads to the antipodal matching of Lorentz symmetries at past and future null infinity in the limit to spatial infinity. We will then show that, assuming a certain continuity condition on the Weyl tensor (\cref{eq:beta-component}) on the boundaries \(\nulls^\pm\) of $\ms{C}$, the Lorentz charges on limiting cross-sections of future (past) null infinity match the Lorentz charges at spatial infinity.\footnote{The formulation of charges at spatial infinity has been investigated in \cite{Henneaux:2018cst,Henneaux:2018hdj,Henneaux:2019yax,Prabhu:2019daz}. In \cite{Henneaux:2018cst,Henneaux:2018hdj,Henneaux:2019yax}, the relation of asymptotic symmetries at null infinity with those at spatial infinity has also been studied.} As a consequence of the asymptotic Einstein's equations on $\hyp$, the Lorentz charges are conserved on $\hyp$ which implies that their values on the past and future boundaries \(\nulls^\pm\) of $\ms{C}$ are equal. It then follows that the Lorentz charges at future null infinity match those at past null infinity in the limit to spatial infinity. This result along with the proof of matching of supertranslation symmetries and the associated supermomentum charges in \cite{KP-GR-match} completes the proof of matching of all BMS symmetries and charges.\footnote{It has been claimed in \cite{Capone:2022gme} that the matching can be proven by converting from Bondi-Sachs coordinates (adapted to null infinity) to Beig-Schmidt coordinates (adapted to spatial infinity). These coordinate transformations are valid only if the conformally-completed spacetime is \(C^{>1}\) at \(i^0\) in \emph{both} null and spatial directions. Additionally, their analysis implicitly assumes that the unphysical metric is \(C^{>0}\) in both null and spatial directions. While these assumptions are valid in Kerr-Newman spacetimes, neither of these assumptions have been justified for generic solutions in nonlinear general relativity (see \cref{sec:disc}).}

The rest of this paper is organized as follows. In \cref{sec:AH}, we review the basic setup needed to study the matching of asymptotic symmetries and charges, borrowing heavily from the description given in \cite{KP-GR-match}. In \cref{sec:symms}, we give a brief review of the asymptotic symmetry groups at null and spatial infinity and study various properties of the associated generators that we will need later in our analysis. We then discuss how fixing the conformal freedom near spatial infinity allows us to isolate Lorentz subgroups of the asymptotic symmetry groups at null and spatial infinity and how the antipodal matching of Lorentz symmetries on past and future null infinity (in the limit to spatial infinity) comes about in \cref{sec:fixing-st}. In \cref{sec:matching}, we show that the Lorentz charges on null infinity match those at spatial infinity and that this leads to the matching of the Lorentz charges on past and future null infinity. We conclude in \cref{sec:disc} with a discussion of the assumptions used in our analysis and some open questions. We collect some results that are needed for the calculations in the body of the paper in the appendices.

\hr
We use abstract index notation with indices \(a,b,c,\ldots\) for tensor fields. Quantities in the physical spacetime are denoted with a ``hat'' on top, while the ones in the conformally-completed unphysical spacetime are denoted without a ``hat'' e.g. \(\hat g_{ab}\) denotes the physical metric while \(g_{ab}\) denotes the unphysical metric. Direction-dependent tensor fields are denoted to be $C^{>-1}$ and their limits to $i^{0}$ are represented by a boldface symbol e.g. $\Omega^{\half} C_{abcd}$ is $C^{>-1}$ at $i^{0}$ and \(\dd C_{abcd}(\vec\eta):=\lim_{\to i^{0}} \Omega^{\half} C_{abcd}\) where $\vec \eta$ denotes the spatial directions at $i^0$. Complex conjugates are denoted using ``bars,'' e.g, $\bar{z}$ denotes the complex conjugate of $z$. Moreover, we use $\scri^{\pm}$ to denote future/past null infinity and $\scri$ without any superscripts in contexts that apply both to past and future null infinity. We use ``$\wt $'' to denote the spin-weights of quantities. Finally, we use ``$\equiv$'' to go from tensors with abstract indices to their expressions in some coordinate system and ``$\hateq$'' to denote equality on $\scri$. The rest of our conventions follow those of Wald \cite{Wald-book}. 
\section{Relating past and future null infinity: the construction}
\label{sec:AH}
In this section, we review various elements of the construction that we will use to relate past and future null infinity in asymptotically-flat spacetimes and address ``the matching problem'', that is, the question of how asymptotic symmetries and charges defined on cross-sections of past and future null infinity are related in the limit to spatial infinity. This construction was developed in \cite{KP-GR-match} (see also \cite{KP-EM-match}) and some accompanying results were derived in \cite{Prabhu:2019daz}. Here, we will simply borrow results from these papers without attempting to derive or prove them. The interested reader is referred to the aforementioned papers to fill in the details.

\paragraph*{Asymptotic flatness at null and spatial infinity:} We will work in a class of spacetimes which are asymptotically-flat at null and spatial infinity. This notion of asymptotic flatness is defined using an Ashtekar-Hansen structure \cite{AH, Ash-in-Held}.

\begin{definition}[Ashtekar-Hansen structure \cite{Ash-in-Held}]\label{def:AH}
	A \emph{physical} spacetime \((\hat M, \hat g_{ab})\) has an \emph{Ashtekar-Hansen structure} if there exists another \emph{unphysical} spacetime \((M, g_{ab})\), such that
	\begin{condlist}
		\item \(M\) is \(C^\infty\) everywhere except at a point \(i^0\) where it is \(C^{>1}\),
		\item the metric \(g_{ab}\) is \(C^\infty\) on \(M-i^0\), \(C^0\) along null directions at \(i^0\) and \(C^{>0}\) along spatial directions at \(i^0\),
		\item there is an embedding of \(\hat M\) into \(M\) such that \(\bar J(i^0) = M - \hat M\),
		\item there exists a function \(\Omega\) on \(M\), which is \(C^\infty\) on \(M-i^0\) and \(C^2\) at \(i^0\) so that \(g_{ab} = \Omega^2 \hat g_{ab}\) on \(\hat M\) and
		\begin{condlist}
			\item \(\Omega = 0\) on \(\dot J(i^0)\),
			\item \(\nabla_a \Omega \neq 0\) on \(\scri\),
			\item at \(i^0\), \(\nabla_a \Omega = 0\), \(\nabla_a \nabla_b \Omega = 2 g_{ab}\). \label{cond:Omega-at-i0}
		\end{condlist}
		\item There exists a neighbourhood \(N\) of \(\dot J(i^0)\) such that \((N, g_{ab})\) is  strongly causal and time orientable, and in \(N \inter \hat M\) the physical metric \(\hat g_{ab}\) satisfies the vacuum Einstein equation \(\hat R_{ab} = 0\),
		\item The space of integral curves of \(n^a = g^{ab}\nabla_b \Omega\) on \(\dot J(i^0)\) is diffeomorphic to the space of null directions at \(i^0\), \label{cond:int-curves}
		\item The vector field \(\varpi^{-1} n^a\) is complete on \(\scri\) for any smooth function \(\varpi\) on \(M - i^0\) such that \(\varpi > 0\) on \(\hat M \union \scri\) and \(\nabla_a(\varpi^4 n^a) = 0\) on \(\scri\). \label{cond:complete}
	\end{condlist}
\end{definition}
Here, following \cite{Hawking-Ellis}, we have denoted the causal future of a point $i^{0}$ in $M$ by $J(i^{0})$, its closure by \(\bar J(i^0)\) and its boundary by \(\dot J(i^0) \). Note that null infinity is such that \(\scri =\dot J(i^0) - i^0\). The physical role of the conditions in \cref{def:AH} is detailed in \cite{Ash-in-Held}. In particular, these conditions  imply that the point \(i^0\) is spacelike related to all points in the physical spacetime \(\hat M\), and represents \emph{spatial infinity}. It is worth emphasizing that the metric \(g_{ab}\) is only \(C^{>0}\) along spatial directions approaching \(i^{0}\), that is, the metric is continuous but the metric connection is allowed to have limits which depend on the direction of approach to \(i^0\). This low differentiability structure is essential to allow spacetimes with non-vanishing ADM mass \cite{AH, Ash-in-Held}. Given a particular physical spacetime, the choice of Ashtekar-Hansen structure is ambiguous up to logarithmic translations.\footnote{We will omit describing any details of logarithmic translations in this paper since we will not need them in any calculations. The interested reader is referred to Remark 4.2 of \cite{Prabhu:2019daz}.} We will comment below on how this ambiguity is fixed in our analysis. 

\paragraph*{A review of the geometry of spatial infinity:}For spacetimes satisfying \cref{def:AH}, we have the following structures in the limit to \(i^0\) along spatial directions. Along spatial directions
\be\label{eq:eta-defn}
\dd\eta^a \defn \lim_{\to i^0} \nabla^a \Omh\,,
\ee
determines a \(C^{>-1}\) spatial unit vector field at \(i^0\) representing the spatial directions \(\vec\eta\) at \(i^0\). The space of directions \(\vec\eta\) in \(Ti^0\) is a unit-hyperboloid denoted by \(\hyp\).

Given \(T^{a \ldots}{}_{b \ldots}\), a \(C^{>-1}\) tensor field at \(i^0\) in spatial directions, \(\lim_{\to i^0} T^{a \ldots}{}_{b \ldots} = \dd T^{a \ldots}{}_{b \ldots}(\vec\eta)\) is a smooth tensor field on \(\hyp\). Moreover, the derivatives of \(\dd T^{a \ldots}{}_{b \ldots}(\vec\eta)\)  with respect to the directions \(\vec\eta\) satisfy
\be\label{eq:dd-der-spatial}
    \dd \partial_c \cdots \dd \partial_d \dd T^{a \ldots}{}_{b \ldots}(\vec\eta) = \lim_{\to i^0} \Omh \nabla_c \cdots \Omh \nabla_d T^{a \ldots}{}_{b \ldots}\,,
\ee
where \(\dd \partial_a\) is the derivative with respect to the directions \(\vec \eta\) defined by 
\be\label{eq:dd-derivative-spatial}\begin{split}
	\dd v^c \dd \partial_c \dd T^{a \ldots}{}_{b \ldots}(\vec\eta) & \defn \lim_{\epsilon \to 0} \frac{1}{\epsilon} \big[ \dd T^{a \ldots}{}_{b \ldots}(\vec\eta + \epsilon \vec v) - \dd T^{a \ldots}{}_{b \ldots}(\vec\eta) \big] \quad \text{for all } \dd v^a \in T\hyp \,,\\
	\dd \eta^c \dd \partial_c \dd T^{a \ldots}{}_{b \ldots}(\vec\eta) & \defn 0\,.
\end{split}\ee
The metric \(\dd h_{ab}\) induced on \(\hyp\) by the universal metric \(\dd g_{ab}\) at \(i^0\) satisfies
\be\label{eq:d-eta-h}
    \dd h_{ab} \defn \dd g_{ab} - \dd \eta_a \dd \eta_b = \dd \partial_a \dd \eta_b\,.
\ee
Further, if \(\dd T^{a \ldots}{}_{b \ldots}(\vec\eta)\) is orthogonal to \(\dd\eta^a\) in all its indices then it defines a tensor field \(\dd T^{a \ldots}{}_{b \ldots}\) intrinsic to \(\hyp\). In this case, it follows from \cref{eq:d-eta-h} and \(\dd\partial_c \dd g_{ab} = 0\) (since \(\dd g_{ab}\) is direction-independent at \(i^0\)) that projecting \emph{all} the indices in \cref{eq:dd-der-spatial} using \(\dd h_{ab}\) defines a derivative operator \(\dd D_a\) intrinsic to \(\hyp\) which is also the covariant derivative operator associated with \(\dd h_{ab}\). We also define
\be\label{eq:volume-hyp}
\dd\varepsilon_{abc} \defn - \dd\eta^d \dd\varepsilon_{dabc} \eqsp \dd\varepsilon_{ab} \defn \dd u^c \dd\varepsilon_{cab}\,,
\ee
where \(\dd\varepsilon_{abcd}\) is volume element at \(i^0\) corresponding to the metric \(\dd g_{ab}\), \(\dd\varepsilon_{abc}\) is the induced volume element on \(\hyp\), and \(\dd\varepsilon_{ab}\) is the induced area element on some cross-section of \(\hyp\) with a future-pointing timelike normal \(\dd u^a\) such that \(\dd h_{ab} \dd u^a \dd u^b = -1\). 

Note that \(\hyp\) admits a reflection isometry which can be seen as follows. We introduce coordinates \((\tau, \theta^A)\)  on $\hyp$--- where \(\tau \in (-\infty,\infty)\), and \(\theta^A = (\theta,\phi )\) are the usual spherical coordinates on \(\bb S^2\)--- such that in these coordinates, the metric on \(\hyp\) is
\be\label{eq:h-hyp-tau}
	\dd h_{ab} \equiv - d\tau^2 + \cosh^2\tau (d\theta^2 + \sin^2\theta d\phi^2 )\,.
\ee
Using \(\Upsilon \circ\) to denote the action of the reflection map \(\Upsilon\) on tensor fields on \(\hyp\), we see that
\be\label{eq:reflection-hyp}\begin{split}
    \Upsilon &: \hyp \to \hyp : (\tau, \theta^A) \mapsto (-\tau, -\theta^A) \\
    \text{with } \Upsilon &\circ \dd h_{ab} = \dd h_{ab}\,,
\end{split}\ee
 where \(\theta^A = (\theta, \phi) \mapsto - \theta^A = (\pi - \theta, \phi\pm \pi)\) is the antipodal reflection on \(\bb S^2\); the sign is chosen so that \(\phi \pm \pi \in [0,2\pi)\).

We turn now to studying some properties of the Weyl tensor in the limit to $i^{0}$. Note that the vacuum Einstein equation \(\hat R_{ab} = 0\) can be written as
\be\label{eq:EE}\begin{aligned}
    S_{ab} & = - 2 \Omega^{-1} \nabla_a \nabla_b \Omega + \Omega^{-2} \nabla^{c}\Omega \nabla_{c}\Omega g_{ab}\,, \\
    \Omega^{\half} S_{ab} & = -4 \nabla_a \eta_b + 4 \Omega^{-\half} \lb( g_{ab} - \tfrac{1}{\eta^{2}} \eta_{a} \eta_{b} \rb)\eta_c \eta^c \,,
\end{aligned}\ee
where, as before, \(\eta_a = \nabla_a \Omh\), and \(S_{ab}\) is given by
\be\label{eq:S-defn}
S_{ab} \defn R_{ab} - \tfrac{1}{6} R g_{ab}\,.
\ee
The Bianchi identity \(\nabla_{[a} R_{bc]de} = 0\) on the unphysical Riemann tensor along with \cref{eq:EE} gives the following equations for the unphysical Weyl tensor \(C_{abcd}\) (see \cite{Geroch-asymp} for details).
\begin{subequations}\label{eq:Bianchi-unphys}\begin{align}
	\nabla_{[e} (\Omega^{-1} C_{ab]cd}) = 0 \label{eq:curl-weyl}\,, \\
	\nabla^d C_{abcd} = - \nabla_{[a} S_{b]c}\,. \label{eq:Weyl-S}
\end{align}\end{subequations}

 Since the unphysical metric \(g_{ab}\) is \(C^{>0}\) at \(i^0\), \(\Omh C_{abcd}\) is \(C^{>-1}\) at \(i^0\) \cite{AH}. We denote
\be
\dd{C}_{abcd}(\vec\eta) \defn \lim_{\to i^{0}} \Omega^{\half} C_{abcd}\,.
\ee
The ``electric'' and ``magnetic'' parts of \(\dd C_{abcd}(\vec\eta)\) are, respectively, defined by
\be \label{eq:EB-defn}
    \dd{E}_{ab}(\vec\eta) \defn \dd{C}_{acbd} (\vec\eta) \dd{\eta}^{c} \dd{\eta}^{d} \eqsp \dd{B}_{ab}(\vec\eta) \defn * \dd{C}_{acbd} (\vec\eta) \dd{\eta}^{c}\dd{\eta}^{d}\,.
\ee
where \(*\dd C_{abcd}(\vec\eta) \defn \tfrac{1}{2} \dd \varepsilon_{ab}{}^{ef} \dd C_{efcd} (\vec\eta)\). It follows from the symmetries of the Weyl tensor that both \(\dd E_{ab}(\vec\eta)\) and \(\dd B_{ab}(\vec\eta)\) are orthogonal to \(\dd\eta^a\), symmetric and traceless with the respect to the metric \(\dd h_{ab}\) on $\hyp$. They therefore define smooth tensor fields on \(\hyp\). \cref{eq:curl-weyl} implies that these satisfy 
\be \label{eq:EB-curl}
\dd D_{[a} \dd E_{b]c} =0 \eqsp \dd D_{[a} \dd B_{b]c}=0\,,
\ee
as well as
\be\label{eq:EB-div}
    \dd D^b \dd{E}_{ab} = \dd D^b \dd{B}_{ab}= 0 \,.
\ee

 Note that since \(g_{ab}\) is \(C^{>0}\) at $i^{0}$, \(\Omh S_{ab}\) is also \(C^{>-1}\) at $i^{0}$. We denote \(\dd S_{ab} (\vec\eta) \defn \lim_{\to i^0} \Omh S_{ab}\) and define
\be \label{eq:potentials-defn}
    \dd{E}(\vec\eta) \defn \dd{S}_{ab}(\vec\eta) \dd{\eta}^{a}\dd{\eta}^{b} \eqsp \dd{K}_{ab}(\vec\eta) \defn \dd{h}_a{}^{c} \dd{h}_b{}^{d} \dd{S}_{cd}(\vec\eta) - \dd{h}_{ab} \dd{E}(\vec\eta)\,, 
\ee
which induce the fields \(\dd E\) and \(\dd K_{ab}\) on \(\hyp\). As shown in \cite{AH,Prabhu:2019daz}, multiplying \cref{eq:Weyl-S} by \(\Omega\), taking the limit to \(i^0\), and using \cref{eq:EB-curl}, we get
\be\label{eq:h-eta-S}
    \dd h_a{}^b \dd \eta^c \dd S_{bc}(\vec\eta) = \dd D_a \dd E\,,
\ee
and
\be \label{eq:EB-potentials}
    \dd{E}_{ab} = -\tfrac{1}{4} (\dd D_{a}\dd D_{b}\dd{E} +  \dd h_{ab} \dd{E}) \eqsp \dd{B}_{ab} = -\tfrac{1}{4}\dd{\varepsilon}_{cda}\dd D^{c}\dd{K}^{d}{}_{b}\,.
\ee
Hence, \(\dd E\) is a scalar potential for \(\dd E_{ab}\) while \(\dd K_{ab}\) is a tensor potential for \(\dd B_{ab}\).\footnote{Since \(\dd B_{ab}\) is curl-free (\cref{eq:EB-curl}), there also exists a scalar potential for \(\dd B_{ab}\) (see, e.g, \cite{AH} and Appendix~B of \cite{Prabhu:2019daz}). However this scalar potential cannot be obtained as the limit of some tensor field in spacetime.} The potentials \(\dd E\) and \(\dd K_{ab}\) are not free fields on \(\hyp\) and are governed by equations which will not concern us here but may be found in \cite{Prabhu:2019daz}.

To define the charge for asymptotic Lorentz symmetries at spatial infinity, which will be part of our analysis in this paper, we will also need access to a ``subleading'' piece of the magnetic part of Weyl tensor for which one has to restrict to a class of spacetimes where $\dd{B}_{ab} =0$. While it is possible to define  Lorentz charges in cases where $\dd{B}_{ab} \neq 0$ (see \cite{CD,Prabhu:2019daz}), that formula is significantly more complicated and therefore we will not analyze those cases here. We point out that the condition \(\dd B_{ab} = 0\) is satisfied (at least) in any asymptotically-flat spacetime which is \emph{either} stationary \emph{or} axisymmetric \cite{B-zero}. Having set \(\dd B_{ab} = 0\), we then require that 
\be \label{eq:beta-defn}
    \dd{\beta}_{ab} \defn \lim_{\to i^{0}} * C_{acbd} \eta^{c} \eta^{d}\,,
\ee
 exists as a \(C^{>-1}\) tensor field at \(i^0\). This defines for us the aforementioned subleading magnetic field. It follows from \cref{eq:beta-defn} that \(\dd\beta_{ab}\) is tangent to \(\hyp\), symmetric and traceless with the respect to \(\dd h_{ab}\). In what follows, we will also need the equations of motion for \(\dd\beta_{ab}\). Our main calculations will be performed in a conformal frame where $\dd{K}_{ab}=0$ (see \cref{rem:shear-fall-off}) and in this frame, these equations are given by \cite{Beig,AH}
\begin{subequations}
\be
    \dd D^b \dd\beta_{ab} =  0 \,.
     \label{eq:div-beta-2}
     \ee
     \be
     \dd{D}^{2} \dd\beta_{a b}-2 \dd \beta_{a b}=-\dd\varepsilon_{c d(a} \dd{E}^{c}{ }_{b)} \dd{D}^{d}\dd{E} \,. \label{eq:wave-eqn-beta-2}
     \ee
\end{subequations}
\begin{remark}[Conformal transformations of the asymptotic fields]\label{rem:conf-GR-fields}
 \cref{def:AH} implies that the allowed conformal freedom \(\Omega \mapsto \omega\Omega\) is such that $\omega>0$ is a positive function which is smooth on $M-i^{0}$, $C^{>0}$ at $i^{0}$ and satisfies $\omega|_{i^{0}}=1$. This implies that along spatial directions that limit to $i^{0}$, we can write
    \be \label{eq:conf-freedom}
    \omega = 1 + \Omega^{\half} \alpha\,,
    \ee    
    where $\alpha$ is $C^{>-1}$ at $i^{0}$. We denote $\dd{\alpha} = \lim_{\to i^{0}} \alpha$. It can then be shown that \(\dd E_{ab}\), \(\dd B_{ab}\) and \(\dd E\) are conformally invariant while
\be\label{eq:conf-K}
    \dd K_{ab} \mapsto \dd K_{ab} - 2 (\dd D_a \dd D_b \dd\alpha + \dd h_{ab}\dd\alpha)\,.
\ee 
\end{remark}

\paragraph*{A review of the geometry of null infinity:}We now introduce some quantities at null infinity that we will need in our analysis. We denote
\begin{subequations}
\be\label{eq:Phi-defn}
    \Phi \defn \tfrac{1}{4} \nabla_a n^a\vert_\scri\,,
    \ee
    \be \label{eq:Phi-at-i0}
     \Phi\vert_{i^0} = 2 \,,
\ee
\end{subequations}
where the second equality follows from \cref{cond:Omega-at-i0}. Under conformal transformations (\cref{rem:conf-GR-fields}), 
\be \label{eq:Phi-conf-tr}
\Phi \to \omega^{-1} (\Phi + \pounds_{n} \ln \omega) \,.
\ee
 Since \(S_{ab}\) is smooth at \(\scri\) by the conditions in \cref{def:AH}, \cref{eq:EE} implies
\be\label{eq:n-Phi}
    \lim_{\to \scri} \Omega^{-1}n^a n_a = 2 \Phi \eqsp \nabla_a n_b \hateq  \Phi g_{ab} \,,
\ee 
that is, the vector field \(n^a\) is a null geodesic generator of \(\scri^\pm \cong \bb R \times \bb S^2\) which is future pointing on $\scri^{+}$ and past pointing on $\scri^{-}$.
Further, we denote the pullback of \(g_{ab}\) to \(\scri\) by \(q_{ab}\). This defines a degenerate metric on \(\scri\) with \(q_{ab} n^b = 0\). It is convenient to introduce a foliation of \(\scri\) by a family of cross-sections diffeomorphic to \(\bb S^2\). The pullback of \(q_{ab}\) to any cross-section \(S\) defines a Riemannian metric on \(S\). Then, for any choice of foliation, there is a unique \emph{auxiliary normal} vector field \(l^a\) at \(\scri\) such that
\be\label{eq:l-props}
    l^a l_a \hateq 0 \eqsp l^a n_a \hateq -1 \eqsp q_{ab}l^b = 0\,.
\ee
 We further have
\be\label{eq:null-fields}
    q_{ab} \hateq g_{ab} + 2 n_{(a} l_{b)} \eqsp \varepsilon_{abc} \hateq l^d \varepsilon_{dabc} \eqsp \varepsilon_{ab} \hateq n^c \varepsilon_{cab}
\ee
where \(\varepsilon_{abc}\) defines a volume element on \(\scri\) and \(\varepsilon_{ab}\) is the area element on any cross-section of the foliation. Evaluating the pullback of \(\Lie_n g_{ab}\) and using \cref{eq:n-Phi}, we have on \(\scri\)
\be\label{eq:Lie-n-q}
    \Lie_n q_{ab} \hateq 2 \Phi q_{ab} \,,
\ee
that is, \(\Phi\) measures the expansion of the chosen cross-sections of \(\scri\) along the null generator \(n^a\) while their shear and twist vanish identically.
We also define
\be\label{eq:tau-defn}
    \tau_a \defn q_a{}^c n^b \nabla_b l_c \,,
\ee
which satisfies \(n^b \nabla_b l_a \hateq \tau_a - \Phi l_a\). We see that \(\tau_a\) represents the change in the direction of \(l_a\) along the null generators of \(n^a\). The shear of the auxiliary normal \(l^a\) on the cross-sections \(S\) of the foliation is defined by
\be\label{eq:sigma-defn}
    \sigma_{ab} \defn \STF \nabla_a l_b \,,
\ee
where $\STF$ denotes the operation of taking the symmetric trace-free projection of a tensor onto a cross-section. The twist \(\varepsilon^{ab}\nabla_a l_b\) vanishes since \(l_a\) is normal to the cross-sections while the expansion of $l^{a}$ is given by
\be \label{eq:expansion-defn}
\vartheta(l^{a}) := q^{ab} \nabla_{a} l_{b}\,.
\ee
For any smooth \(v_a\) satisfying \(n^a v_a \hateq l^a v_a \hateq 0\), we define the derivative \(\ms D_a\) on the cross-sections by
\be
    \ms D_a v_b \defn q_a{}^c q_b{}^d \nabla_c v_d\,.
\ee
It is easily verified that \(\ms D_a \varepsilon_{bc} \hateq 0\) and \(\ms D_a q_{bc} \hateq 0\).

In this paper, we will work in a class of spacetimes where the peeling theorem holds. It follows then that \(C_{abcd} = 0\) at \(\scri\), and thus \(\Omega^{-1}C_{abcd}\) admits a limit to \(\scri\) (see, e.g, Theorem~11 of \cite{Geroch-asymp}). In any choice of foliation of \(\scri\) we define the fields
\begin{subequations}\label{eq:weyl-defn}\begin{align}
    \mc R_{ab} & \defn (\Omega^{-1}C_{cdef}) q_a{}^c n^d q_b{}^e n^f \eqsp & \mc S_a & \defn (\Omega^{-1}C_{cdef}) l^c n^d q_a{}^e n^f \\
    \mc P & \defn (\Omega^{-1}C_{cdef}) l^c n^d l^e n^f \eqsp & \mc P^* & \defn \tfrac{1}{2} (\Omega^{-1}C_{cdef}) l^c n^d \varepsilon^{ef} \label{eq:P-defn2}\\
    \mc J_a & \defn (\Omega^{-1}C_{cdef}) n^c l^d q_a{}^e l^f \eqsp & \mc I_{ab} & \defn (\Omega^{-1}C_{cdef}) q_a{}^c l^d q_b{}^e l^f \label{eq:non-peeled}
\end{align}\end{subequations}
These tensors are all orthogonal to \(n^a\) and \(l^a\) in all indices and therefore can be taken to be tensor fields on the cross-sections of the chosen foliation of \(\scri\). Relations of these tensor fields to Weyl scalars on null infinity may be found in Appendix.~A of \cite{KP-GR-match,GPS}.\footnote{Expressed in conformal Bondi-Sachs coordinates, $\mc{P}$ and $\mc{J}_{a}$ are related to the Bondi mass aspect, $M$, and the angular momentum aspect $N_{A}$; these relations may be found in Eq.~6.10 of \cite{GPS}.}
For the fields defined in \cref{eq:weyl-defn}, \cref{eq:curl-weyl} implies the following evolution equations along \(\scri\)
\begin{subequations}\label{eq:weyl-evol}\begin{align}
    (\Lie_n + 2\Phi) \mc S_a & = (\ms D^b + \tau^b) \mc R_{ab}\,, \\
     (\Lie_n + 3\Phi) \mc P~ & = (\ms D^a + 2 \tau^a) \mc S_a - \sigma^{ab} \mc R_{ab} \label{eq:evol-P}\,, \\
     (\Lie_n + 3\Phi) \mc P^* & = - \varepsilon^{ab}(\ms D_a + 2 \tau_a) \mc S_b + \varepsilon_b{}^c \sigma^{ab} \mc R_{ac}\,, \\
    (\Lie_n + 2 \Phi) \mc J_a & = \tfrac{1}{2}(\ms D_b + 3\tau_b ) (q_a{}^b \mc P - \varepsilon_a{}^b \mc P^*) - 2 \sigma_a{}^b \mc S_b \label{eq:evol-J}\,, \\
    (\Lie_n + \Phi) \mc I_{ab} & =   (q_a{}^c q_b{}^d - \tfrac{1}{2} q_{ab} q^{cd} )  (\ms D_c + 4 \tau_c) \mc J_d - \tfrac{3}{2} \sigma_{ac} (q_b{}^c \mc P - \varepsilon_b{}^c \mc P^*)\,.
\end{align}\end{subequations}

Finally, the News tensor is defined by
\be\label{eq:News-defn}
    N_{ab} \defn 2 (\Lie_n - \Phi ) \sigma_{ab}\,,
\ee
which satisfies \(N_{ab}n^b \hateq 0\), \(N_{ab}q^{ab} \hateq 0\) and is conformally-invariant on \(\scri\). The News tensor is related to \(S_{ab}\) (\cref{eq:S-defn}) by
\be\label{eq:News-Ric}
    N_{ab} \hateq  \STF \lb[ S_{ab} - 2\Phi \sigma_{ab} + 2 (\ms D_a \tau_b + \tau_a \tau_b) \rb] \,,
\ee
and to the Weyl tensor on \(\scri\) by (from \cref{eq:Weyl-S})
\begin{subequations}
\be
 \mc R_{ab} \hateq \tfrac{1}{2} \Lie_n N_{ab} \label{eq:weyl-News-1} 
 \ee \be
  \mc S_a \hateq \tfrac{1}{2} \ms D^b N_{ab} \label{eq:weyl-News-2}\,.
 \ee
    \end{subequations}

\begin{figure}[h!]
	\centering
	\includegraphics[width=0.3\textwidth]{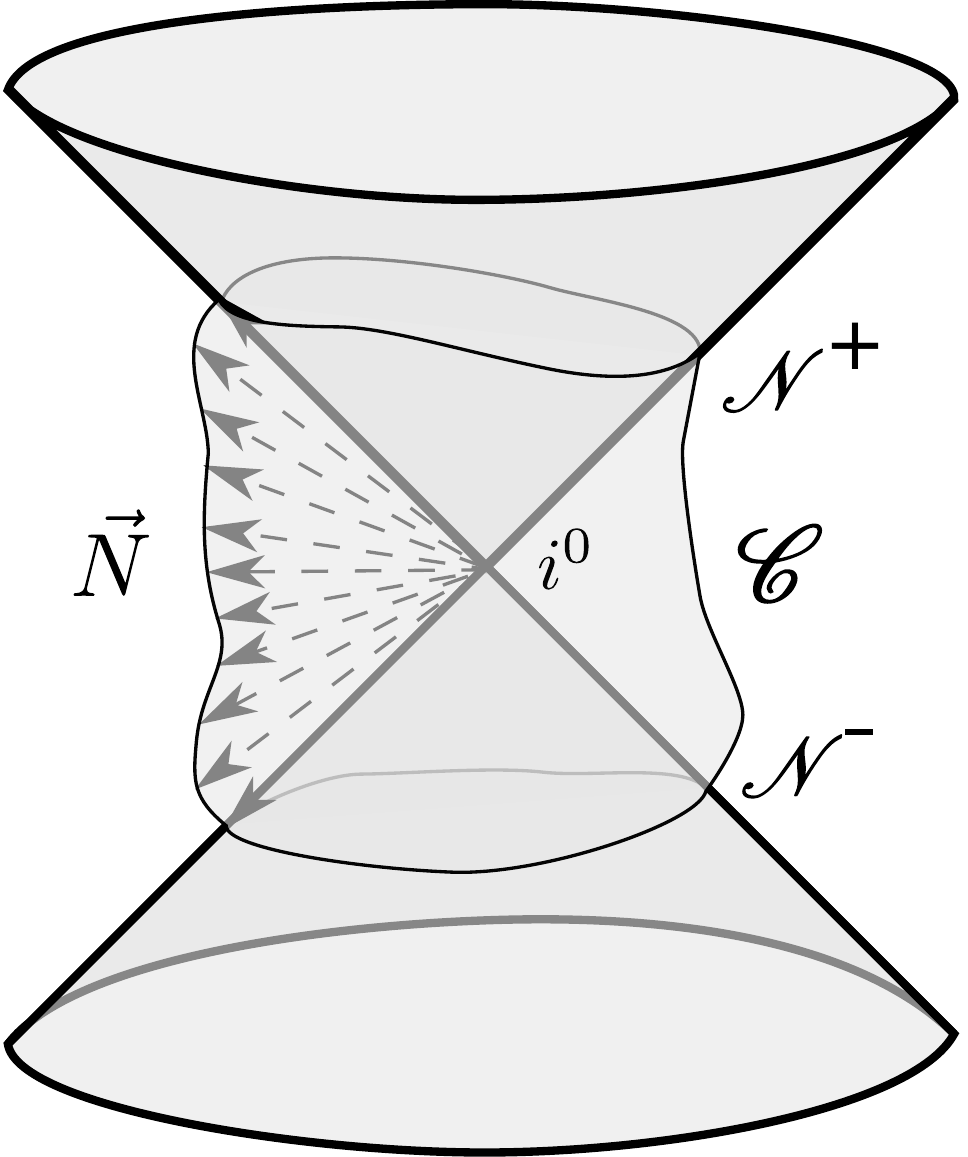}
	\caption{The space \(\cyl\) of null and spatial directions \(\vec N\) at \(i^0\). The boundaries \(\nulls^\pm \cong \bb S^2\), diffeomorphic to the space of generators of \(\scri^\pm\) respectively, represent the space of null directions. \(\cyl \setminus \nulls^\pm\) is the space of rescaled spatial directions conformally diffeomorphic to the unit-hyperboloid \(\hyp\). \(\cyl\) depends on the choice of the rescaling function \(\Sigma\) (defined below) and need not be a cylinder of unit radius in \(Ti^0\) --- a ``wiggly'' cylinder has been drawn here to emphasize this.} 
    \label{fig:cylinder}
\end{figure}

\paragraph*{The space $\ms{C}$ of null and spatial directions at $i^{0}$:}  As detailed in \cite{KP-GR-match}, to study limits of quantities defined at null infinity to spatial infinity, one needs to rescale $n^{a}$ to obtain a set of ``good'' (non-vanishing) null directions at $i^{0}$ and, in addition, conformally complete $\hyp$ into a cylinder, denoted by $\mathscr{C}$. The boundaries of $\ms{C}$, denoted by $\nulls^{\pm}$, correspond to the space of (rescaled) null directions at $i^{0}$ (see \cref{fig:cylinder} for an illustration) that are antipodally mapped onto each other by the reflection map (\cref{eq:reflection-hyp}). To carry out this rescaling of directions, in a neighborhood of $i^{0}$ in $M$ (from hereon in, we use $M$ to denote such a neighborhood unless otherwise specified), one defines 
\be\label{eq:N-defn}
N^{a}:= \frac{1}{2} \Sigma n^{a} = \frac{1}{2} \Sigma \nabla^{a} \Omega \,,
\ee
where $\Sigma$, called the rescaling function, satisfies the properties listed below.

\begin{definition}[Rescaling function \(\Sigma\)]
\label{def:Sigma}
We take \(\Sigma\) to be a function in $M$  such that
\begin{condlist}
    \item \(\Sigma^{-1} > 0\) is smooth on \(M - i^0\)
    \item \(\Sigma^{-1}\) is \(C^{>0}\) at \(i^0\) in both null and spatial directions,
	\item \(\Sigma^{-1}\vert_{i^0} = 0\), \(\lim_{\to i^0} \nabla_a \Sigma^{-1} \neq 0\) and \label{cond:Sigma-i0}
    \item \(\Sigma \Lie_n \Sigma^{-1} = 2\) at \(i^0\) and on \(\scri\) \label{cond:Sigma-choice}
\end{condlist}
\end{definition}
Note that $\Sigma$ is not uniquely defined and the freedom in picking a $\Sigma$ is detailed in Remark 2.3 of \cite{KP-GR-match}. Note also that $N^{a}$ is $C^{>-1}$ at $i^{0}$ and $\dd{N}^{a}= \lim_{\to i^{0}} N^{a} \neq 0$ along both null and spatial directions. 
\\

Since \(\Sigma^{-1}\vert_{i^0} = 0\) and \(\Sigma^{-1}\) is \(C^{>0}\), there exists a function \(\dd\Sigma(\vec\eta)\), which is \(C^{>-1}\) along spatial directions, such that
\be\label{eq:dd-Sigma}
    \dd\Sigma^{-1}(\vec\eta) = \lim_{\to i^0} (\Omh \Sigma)^{-1} \,.
\ee

We also define a rescaled auxiliary normal \(L^a\) in $M$ by
\be \label{eq:Lamodified}
L^{a} := - \nabla^{a} \Sigma^{-1} + \frac{1}{2} N^{a} \nabla^{b} \Sigma^{-1} \nabla_{b} \Sigma^{-1} - \frac{1}{2} \Omega \Sigma \bar{L}^{b} \nabla^{a} \nabla_{b} \Sigma^{-1}\,,
\ee
where
\be \label{eq:barLa}
\bar{L}^{a} := - \nabla^{a} \Sigma^{-1} + \frac{1}{2} N^{a} \nabla_{b} \Sigma^{-1} \nabla_{b} \Sigma^{-1}\,.
\ee
Our $L^{a}$, we note, is different from the expression for $L^{a}$ (here denoted as $\bar{L}^{a}$) in \cite{KP-GR-match} where the last term was absent.\footnote{\label{footnote:differntl's}Note that the expression for $\dd{\bar{L}}^{a}$ in Eq.~2.33 of \cite{KP-GR-match} was erroneously written. The corrected version of that expression is $\dd{\bar{L}}^{a}=-\dd{h}^{a}{}_{b}\dd{D}^{b} \dd{\Sigma}^{-1} + \dd{\eta}^{a} (\frac{1}{2} \dd{\Sigma} \dd{h}^{bc}\dd{D}_{b} \dd{\Sigma}^{-1} \dd{D}_{c} \dd{\Sigma}^{-1} -\frac{1}{2} \dd{\Sigma}^{-1})$. However, since this correction does not effect $\dd{h}^{a}{}_{b} \dd{\bar{L}}^{b}$, it does not effect any of the conclusions in \cite{KP-GR-match}. It can also be shown explicitly that $\lim_{\to \nulls^\pm}\dd{\Sigma}^{-1} \dd{h}^{a}{}_{b} \dd{L}^{b}=\lim_{\to \nulls^\pm}\,\dd{\Sigma}^{-1} \dd{h}^{a}{}_{b} \dd{\bar{L}}^{b}$ and so the proof of matching of supertranslation charges in \cite{KP-GR-match} goes through unchanged with the choice of $L^{a}$ made in this paper.}
Note also that $L^{a}$ is \(C^{>-1}\) at \(i^0\) and \(\lim_{\to i^0} L^a \neq 0\) in both null and spatial directions. Further, using \cref{eq:N-defn,cond:Sigma-choice}, we have
\be\label{eq:NL-LL}
    N^a L_a\hateq -1 \eqsp L^a L_a \hateq 0\,.
\ee
The pullback of \(L_a\) to $\scri$ equals the pullback of \(- \nabla_a \Sigma^{-1}\) and therefore \(L^a\) defines a rescaled auxiliary normal to a foliation of \(\scri\) by a family of cross-sections \(S_\Sigma\) with \(\Sigma^{-1} = \text{constant}\). It follows from \cref{def:Sigma} and \cref{cond:int-curves} that the limiting cross-section \(S_\Sigma\) as \(\Sigma^{-1} \to 0\), is diffeomorphic to \(\nulls^\pm\). The auxiliary normal to this foliation, $l^{a}$, satisfying \cref{eq:l-props}, is obtained by
\be\label{eq:l-defn}
    l^a \defn \tfrac{1}{2} \Sigma L^a\,,
\ee
which we also take to define our choice of extension of $l^{a}$ into $M$. In the foliation of $S_{\Sigma}$ cross-sections, we have (using \cref{eq:N-defn,eq:NL-LL})
\begin{subequations}
\be
    N^a\vert_\scri \equiv \partial_{\Sigma^{-1}} \,,
    \ee
    \be
     n^a\vert_\scri \equiv 2 \Sigma^{-1} \partial_{\Sigma^{-1}} \label{eq:N-n-Sigma}  \,. 
\ee
\end{subequations}
We turn now to the conformal completion of $\hyp$. Let \(\dd\Sigma\) be the function induced on \(\hyp\) by \(\dd\Sigma(\vec\eta)\) (defined in \cref{eq:dd-Sigma}). Let \((\tilde\hyp, \tilde{\dd h}_{ab}) \) be a conformal-completion of \((\hyp, \dd h_{ab})\) with the metric  \(\tilde{\dd h}_{ab} := \dd\Sigma^2 \dd h_{ab}\). Then there exists a diffeomorphism from \(\tilde\hyp\) onto \(\cyl\) (see Eq.~B.16 of \cite{KP-EM-match}) such that \(\hyp\) is mapped onto \(\cyl \setminus \nulls^\pm\) and \(\dd\Sigma\), as a function on \(\cyl \setminus \nulls^\pm\), extends smoothly to the boundaries \(\nulls^\pm\) where
\be
    \dd\Sigma\vert_{\nulls^\pm} = 0 \,.
\ee
Note that we will implicitly use this diffeomorphism to treat fields defined on $\hyp$ as fields on $\cyl$ throughout this paper. Note also that the rescaled metric
\be\label{eq:rescaled-q}
    \tilde q_{ab} \defn \Sigma^2 q_{ab}\,,
\ee
on \(S_\Sigma\) is such that as \(\Sigma^{-1}\to0\), \(\lim_{\to i^0} \tilde q_{ab}(\vec N)\) exists along null directions \(\vec N\) and defines a direction-dependent Riemannian metric \(\tilde{\dd q}_{ab}\) on the space of null directions \(\nulls^\pm\). Moreover, this metric coincides with the metric induced on \(\nulls^\pm\) by \(\tilde{\dd h}_{ab}\) on \(\cyl\), that is, 
\be
    \tilde{\dd q}_{ab} = \lim_{\to \nulls^\pm} (\tilde{\dd h}_{ab} + \dd D_a \dd\Sigma \dd D_b \dd\Sigma) \,.
\ee
Similarly, we have the rescaled area element 
\be \label{eq:rescaledareaelement}
\tilde\varepsilon_{ab} \defn \Sigma^2 \varepsilon_{ab} \,,
\ee
 on the foliation \(S_\Sigma\). This induces an area element \(\tilde{\dd\varepsilon}_{ab}\) on \(\nulls^\pm\) such that 
\be\label{eq:nulls-area}
    \tilde{\dd\varepsilon}_{ab} = \lim_{\to \nulls^\pm} \dd U^c \tilde{\dd\varepsilon}_{cab} = \lim_{\to \nulls^\pm} \pm \dd{\Sigma}^{2}\, \dd{\varepsilon}_{ab} \,,
\ee
where $\dd{U}^{a} := \dd{h}^{a}{}_{b} \dd{L}^{b}$ and \(\tilde{\dd\varepsilon}_{abc} \defn \dd\Sigma^3 \dd\varepsilon_{abc}\) is the volume element on \(\cyl\) defined by the metric \(\tilde{\dd h}_{ab} = \dd\Sigma^2 \dd h_{ab}\). Note also that
\be \label{eq:relbetnormals}
\lim _{\rightarrow \mathscr{N} \pm} \dd{\Sigma^{-1}} \dd{U}^{a}=\pm \lim _{\rightarrow \mathscr{N} \pm} \dd{\Sigma^{-2}} \dd{u}^{a} \neq 0 \,.
\ee

\paragraph*{Null-regular spacetimes at $i^{0}$:}\label{para:null-reg} 
In \cite{KP-GR-match}, it was shown that the spacetimes in which the supertranslations charges on $\scri^-$ and $\scri^+$ match in the limit to $i^{0}$ are spacetimes with an Ashtekar-Hansen structure (\cref{def:AH}) where 
\begin{enumerate}
    \item the rescaled quantity
    \be\label{eq:P-falloff}
    \Sigma^{-3} \Omega^{-1} C_{abcd}l^a n^b l^c n^d \text{ is } C^{>-1} \text{ in \emph{both} null and spatial directions at } i^0
    \ee
    \item in the limit to \(i^0\) along each null generator of \(\scri\)
    \be\label{eq:N-R-falloff}
     N_{ab} = O(\Sigma^{- (1+\epsilon)}) \eqsp \mc R_{ab} = O(\Sigma^{-(1+\epsilon)}) \text{ as } \Sigma^{-1} \to 0 \text{ along } \scri
    \ee
\end{enumerate}
for the vector field \(l^a\) (defined by \cref{eq:Lamodified,eq:l-defn}; see \cref{footnote:differntl's} as well). Such spacetimes are called \emph{null-regular} at \(i^0\). It was also shown that these spacetimes satisfy the property that $\dd{E}_{ab}$ is even under the reflection isometry given in \cref{eq:reflection-hyp}. 
Further, it was shown in \cite{Prabhu:2019daz} (in Eq. 4.31 and the discussion around it) that in such spacetimes, one can use logarithmic translations to set $\dd{E}$ to be reflection-even. This then removes the ambiguity in the Ashtekar-Hansen structure referred to below \cref{def:AH}. Throughout this paper, we will always work in these null-regular spacetimes. We will later supplement this with the condition given in \cref{eq:shearfalloff} and additionally assume that \cref{eq:beta-component} is continuous at $\nulls^{\pm}$. Note that in our analysis, we do not require that the News tensor vanish in any open region of $\scri$ and therefore do \emph{not} impose that our spacetimes be stationary (see pp. 53-54 of \cite{Geroch-asymp} for a proof of the statement that if a spacetime is stationary in a neighborhood of some portion of $\scri$, labeled $\Delta\scri$, then $N_{ab} \hateq 0$ on $\Delta \scri$).
\paragraph*{Choice of conformal frame:} \label{para:conf-choice}
One can use the conformal freedom $\Omega  \to  \omega\Omega$ discussed in \cref{rem:conf-GR-fields} and the corresponding change in $\Phi$, given in \cref{eq:Phi-conf-tr}, to go to a conformal frame where $\Phi=2$ not just at $i^{0}$ but in a neighborhood of $i^{0}$. The appropriate $\omega$ can be picked by solving the following ordinary differential equation $\Lie_{n} \ln \omega = 2 \omega - \Phi$ (for some initial $\Phi$) in a neighborhood of $i^{0}$. This was done, e.g in \cite{Ash-Mag-Ash,HL-GR-matching}. In what follows, we will also work in a conformal frame where this is true and so all our subsequent calculations will be performed assuming that we work on a portion of $\scri$ that is in a neighborhood of $i^{0}$  where $\Phi=2$. Since the asymptotic charges at both null and spatial infinity are conformally invariant (see, e.g, \cite{GPS,Prabhu:2019daz}), making this choice to prove matching of asymptotic charges entails no loss of generality. Note also that after having picked this conformal frame, one still has the residual conformal freedom given by $\Omega \to \omega \Omega$ where $\Lie_{n} \omega \hateq 0$. We will restrict this freedom further in \cref{sec:fixing-st}.
\paragraph*{Choice of rescaling function:} Note that since asymptotic charges at both null and spatial infinity are independent of the choice of rescaling function, we can use any choice of rescaling function to study the matching of asymptotic charges. A particularly convenient choice is one where the metric on cross-sections of $\scri$ (in a neighborhood of $i^{0}$) is $q_{ab} \hateq \Sigma^{-2} s_{ab}$ where $s_{ab}$ is (a constant multiple of) the unit round sphere metric. This choice can always be made when $\Phi=2$ (see e.g, Appendix.~B of \cite{KP-GR-match}) and in the rest of our analysis we will work with this choice.

\paragraph*{Choice of foliation:} As in \cite{KP-GR-match}, in the rest of this paper, we will work in a context where $\scri$ is foliated by $\Sigma^{-1}=\text{constant}$ cross-sections, $S_{\Sigma}$ . This implies that
\be \label{eq:divsigma=0}
\ms{D}_{a} \Sigma^{-1} \hateq 0 \,,
\ee
on any cross-section of the foliation. It was shown in \cite{KP-GR-match} that this choice can be made in any conformal frame and that using \cref{eq:divsigma=0}, this implies that $\tau_{a} \hateq 0$ (see Eq. 2.31 and footnote 5 of \cite{KP-GR-match}). Therefore, $\tau_{a}\hateq 0$ in the rest of our analysis.\\
\paragraph*{Limits of integrals to $\scri$ and $\hyp$:} \label{para:surfaces}
In this paper, we will need to consider limits of certain integrated quantities to cross-sections of $\scri$ and $\hyp$ (see \cref{sec:matching}). In these cases, the limits to cross-sections of $\scri$ will be taken along a sequence of null hypersurfaces, that exists in a neighborhood of $i^{0}$ in the unphysical spacetime. Each of these null surfaces is foliated by constant $\Omega$ spheres $\mc{S'}$, that limit to cross-sections $S_{\Sigma}$ of $\scri$. These null surfaces are taken to be generated by an affine, null vector field, denoted by $K^{a}$ (defined in \cref{eq:startingpoint-1}). This vector field is such that $\lim_{\to i^{0}} \Omega^{\half} K^{a}$ is direction-dependent. Moreover, we require that the null normal $K_{a}$ be such that $\lim_{\to i^{0}} K_{a}$ is direction-dependent. Note that this difference in scaling between the null generator and null normal uses the fact that on a null surface they can be scaled arbitrarily with respect to each other. The limit to cross-sections of $\hyp$ is taken along spacelike hypersurfaces that go to spatial infinity. These surfaces are foliated by spheres, $\mc{S}'$, that limit to cross-sections of $\hyp$. \\

 This completes our review of the construction needed for the calculations in this paper.
\section{Asymptotic symmetries at spatial infinity}
\label{sec:symms}

\subsection{Behaviour of the BMS symmetries at \(i^0\)}
\label{sec:bms-gen}

In this section, we derive the behavior of BMS symmetries in the limit to $\nulls^{\pm}$ along $\scri^{\pm}$. We start by giving a brief review of BMS symmetries (see, e.g, \cite{GPS} for a more detailed discussion).

 BMS symmetries at null infinity are defined by diffeomorphisms that preserve the universal structure at $\scri$ (that is, the structure common to all physical spacetimes that satisfy \cref{def:AH}). This universal structure  is given by the equivalence class $[n^{a}, q_{ab}]$ with $(n^{a}, q_{ab}) \sim (\omega^{-1} n^{a}, \omega^{2} q_{ab})$, where $\omega$ is a positive function which is smooth on $M-i^{0}$, $C^{>0}$ at $i^{0}$ and satisfies $\omega|_{i^{0}}=1$. The diffeomorphisms on \(\scri\) which preserve this universal structure are generated by vector fields $\xi^{a}$, in the physical spacetime, which extend smoothly to $\scri$, are tangent to $\scri$ and satisfy 
\be\label{eq:symm-null}
    \Lie_\xi n^a \hateq - \alpha_{(\xi)} n^a \eqsp \Lie_\xi q_{ab} \hateq 2 \alpha_{(\xi)} q_{ab}\,,
\ee
for some function \(\alpha_{(\xi)}\) which depends on \(\xi^a\), is smooth on \(\scri\), \(C^{>0}\) in spatial directions at \(i^0\) and satisfies \(\alpha_{(\xi)} \vert_{i^0} = 0\) (which follows from the fact that  \(\omega \vert_{i^0} = 1\) ). 

Since $\xi^{a}$ is tangent to $\scri$, we can write 
\be \label{eq:BMSsymmonscri}
    \xi^a \hateq \beta n^a + q^{a}{}_{b} X^b \,,
\ee

where $\beta:= f - l_{a} X^{a}$. \cref{eq:symm-null} then gives\footnote{Recall that we have specialized here to $\Phi=2$ and $\tau_{a} \hateq 0$. The corresponding expressions in arbitrary conformal frames and arbitrary foliations of $\scri$ may be found in Appendix.~A of \cite{GPS}.}
\begin{subequations}
   \be \label{eq:bms-cond-1} (\Lie_n - 4)\, q^{b}{}_{a} X_b \hateq 0 \,,
   \ee
   \be   \label{eq:bms-cond-2}
    (q_a{}^c q_b{}^d - \tfrac{1}{2} q_{ab} q^{cd} ) \ms D_{(c}\, q^{e}{}_{d)} X_{e}  \hateq 0\,, 
   \ee
   \be \label{eq:bms-cond-3}
    \alpha_{(\xi)} \hateq \Lie_n\beta \hateq 2 \beta + \tfrac{1}{2} \ms D_a (q^{a}{}_{b} X^a)\,, \\
    \implies (\Lie_n -2) \beta \hateq \tfrac{1}{2} \ms D_a (q^{a}{}_{b} X^a)\,.
    \ee
\end{subequations}
 BMS symmetries on $\scri$ are parametrized by $(f,  X^a)$ that satisfy these conditions. It can be shown (see, e.g \cite{GPS,Sachs2,Geroch-asymp}) that symmetries of the form $(f, X^a=0)$ form an infinite dimensional subalgebra, $\mf{s}$, which is a Lie ideal of the BMS algebra, $\mf{b}$. These symmetries are called BMS supertranslations and they satisfy \((\Lie_n - 2)f \hateq 0\) (which follows from \cref{eq:bms-cond-3} with $ X^{a}=0$). Further, one can see from \cref{eq:bms-cond-2} that $q^{a}{_{b} }X^{b}$ satisfies the conformal Killing equation on cross-sections of $\scri$ and since these cross-sections are diffeomorphic to $\mathbb{S}^{2}$, it follows that $q^{a}{}_{b} X^{b}$ are elements of the Lorentz algebra, $\mf{so}(1,3)$. One can show that the Lie bracket of a BMS supertranslation and a Lorentz symmetry is a BMS supertranslation and therefore, the  Lorentz algebra forms a quotient subalgebra of $\mf{b}$. The structure of $\mf{b}$ is therefore that of a semidirect sum,
\be
\mf{b} \cong  \mf{so}(1,3) \ltimes \mf{s}\,.
\ee
 Finally, there is 4-dimensional Lie ideal, $\mf{t}$, of $\mf{s}$ which corresponds to BMS translations. These are BMS supertranslations which satisfy
 \be
 \STF \ms{D}_{a}\ms {D}_{b} f \hateq 0\,.
 \ee
Next, we study the behavior of $\xi^{a}$ in the limit to $\nulls^{\pm}$ along $\scri$ by solving \cref{eq:bms-cond-1,eq:bms-cond-2,eq:bms-cond-3} in a coordinate system adapted to $\scri$ that is well defined in a neighborhood of $i^{0}$ (constructed in Appendix.~B of \cite{KP-EM-match}). In these coordinates, \(q_{ab} \hateq \Sigma^{-2} s_{ab}\) where $s_{ab}$ is the unit round sphere metric, given in stereographic coordinates by (\cref{eq:stereographic-metric})
 \be 
     s_{AB} \equiv 2 P^{-2} d z d\bar z \,,
     \ee
 where $P:= \frac{1+ z \bar{z}}{\sqrt{2}}$. Written in these coordinates,
\be \label{eq:Xisl1}
  q^{a}{}_{b}  X^b \hateq P X \partial_{\bar z} + P \bar X \partial_z\,.
    \ee
    Here $X \wt s =-1$ which is determined by the fact that $q^{a}{}_{b} X^{b}$ is invariant under spin transformations (see \cref{eq:spin-wt-defn} for the definition of spin transformations). Using the definition of $\bar \eth$ in \cref{eq:ethethbaraction}, \cref{eq:bms-cond-2}, written in stereographic coordinates, is the same as
\be
    \bar\eth X \hateq 0\,,
\ee
which, in terms of spherical harmonics on the unit-sphere, implies that \(X\) is  \(\ell=1\). Note also that,
\be \label{eq:div-X-to-eth-ethbar}
    \ms D_a (q^{a}{}_{b} X^{b}) \hateq  (\eth X + \bar\eth\, \bar X )\,.
\ee 
With \(n^a \hateq 2 \Sigma^{-1} \partial_{\Sigma^{-1}}\) (\cref{eq:N-n-Sigma}), \cref{eq:bms-cond-1,eq:bms-cond-3} lead to
\be
    2 \Sigma^{-1} \partial_{\Sigma^{-1}} (\Sigma^{-2} X) - 4 (\Sigma^{-2} X) \hateq 0 \eqsp 2 \Sigma^{-1} \partial_ {\Sigma^{-1}} \beta - 2 \beta \hateq \tfrac{1}{2} \ms D_a (q^{a}{}_{b} X^{b})\,.
\ee
The first equation above implies that $X$ is constant in $\Sigma^{-1}$ and therefore has a well defined limit as \(\Sigma^{-1} \to 0\). It then follows from \cref{eq:div-X-to-eth-ethbar} that $\ms D_a (q^{a}{}_{b} X^{b})$ is also constant in $\Sigma^{-1}$. Further, the solution to the second equation above gives
\be \label{eq:beta-soln}
    \beta \hateq \Sigma^{-1} \beta_0 - \tfrac{1}{4}  \ms D_a (q^{a}{}_{b} X^{b})\,,
\ee
where \(\beta_0\) is a constant in \(\Sigma^{-1}\). Using $\alpha_{(\xi)} \hateq \Lie_{n} \beta$, we then obtain
\be \label{eq:alpha-exp}
 \alpha_{(\xi)} \hateq  2 \Sigma^{-1} \beta_0\,.
 \ee
 Recall that a BMS supertranslation is given by $\xi^{a}|_{X^{a}=0}$ evaluated on $\scri$ and we see that in this case, $\Sigma\, \beta(X^{a}=0)=\Sigma f=\beta_{0}$ has a non-vanishing limit to $\nulls^{\pm}$. As already shown above, $q^{a}{}_{b} X^{b}$ has a limit as $\Sigma^{-1} \to 0$ and we see therefore that a BMS symmetry on $\nulls^{\pm}$ is given by $(\Sigma f, q^{a}{}_{b} X^{b})$ where $q^{a}{}_{b} X^{b}$ is a conformal Killing vector field tangent to $\nulls^{\pm}$. 

\subsection{Spi symmetries on the space of null directions \(\nulls^\pm\)} \label{sec:spi-symm}

In this section, we will study some properties of the asymptotic symmetries at spatial infinity, called \emph{spi} symmetries, that we will need in our analysis. 

Spi symmetries correspond to diffeomorphisms that preserve the universal structure at spatial infinity. The consequences of this were discussed in Sec.~6 of \cite{Prabhu:2019daz} where it was shown that this implies that the generators of these diffeomorphisms, $\xi^{a}$, are such that
   $\lim_{\to i^{0}} \Omega^{-\half} \xi^{a}$ is direction-dependent and $\xi^{a}$ satisfies $\Lie_{\xi} g_{ab}= 4 \Omega^{-\half} \xi^{c}\eta_{c}g_{ab}$ on $\hyp$. Spi symmetries are parametrized on $\hyp$ by $(\dd{f},\dd{X}^{a})$ where
     \be
  \dd{X}^{a} := \lim_{\to i^{0}} \Omega^{-\half} \xi^{a} \eqsp \dd{f}:= \lim_{\to i^{0}} \Omega^{-1} \xi^{a} \eta_{a}\,,
   \ee
     where $\dd{f}$ is a smooth function on $\hyp$ which parameterizes spi supertranslations. The action of a spi supertranslation on the asymptotic fields on $\hyp$ is that of a linearized conformal transformation with $\dd{\alpha} = -2 \dd{f}$ (where $\dd\alpha$ was defined in \cref{rem:conf-GR-fields}). Further, $\dd{X}^{a}$ satisfies $\dd{\eta}_{a} \dd{X}^{a} = \dd{D}_{(a} \dd{X}_{b)} =0$ on $\hyp$ which implies that $\dd{X}^{a}$ is an element of the Lorentz algebra, \(\mf{so}(1,3)\). These symmetries comprise the $\mf{spi}$ algebra which is very similar in its structure to the BMS algebra in that, $\mf{spi} \cong \mf{so}(1,3) \ltimes \mf{s}$; that is, it is given by the semi-direct sum of spi-supertranslations, $\mf{s}$, (which form an infinite dimensional Lie ideal of $\mf{spi}$) with the Lorentz algebra which like in BMS algebra forms a quotient subalgebra of $\mf{spi}$. Finally, the $\mf{spi}$ algebra also has a 4-dimensional subalgebra $\mf{t}$ of spi translations which forms a Lie ideal of $\mf{t}$. These are given by the spi-supertranslations $f$ which satisfy the additional condition
  \be \label{eq:translations-condition}
   \dd{D}_{a} \dd{D}_{b} \dd{f}+\dd{h}_{a b} \dd{f}=0\,.
   \ee
    The limiting behavior of spi supertranslations to $\nulls^{\pm}$ was studied in \cite{KP-GR-match} where it was shown that for spi-supertranslations that match onto BMS-supertranslations at null infinity in the limit to spatial infinity, $\dd{F} \defn \dd{\Sigma} \dd{f}$ has a limit to $\nulls^{\pm}$. Since only these spi supertranslations are relevant for the matching problem we are studying here, we restrict our attention only to them. In this paper, we will also need some properties of $\dd{X}^{a}$, including its limiting behavior to $\nulls^{\pm}$. We turn to deriving that next.
   \subsubsection*{Lorentz symmetries on \(\hyp\)} \label{sec:LorentzsymmonH}
 To study the limiting behaviour of $\dd{X}^{a}$, we explicitly solve \(\dd D_{(a}\dd X_{b)} =0 \) and analyze the behavior of the solutions to this equation in the limit to $\nulls^{\pm}$. We use coordinates  $(\alpha, z,\bar{z})$ on $\hyp$ (see Appendix.~B of \cite{KP-GR-match} for details), in which the metric on $\hyp$, $\dd{h}_{ab}$, has the following form
 \be \label{eq:hyp-metric}
 \dd{h}_{a b} \equiv-\frac{1}{\left(1-\alpha^{2}\right)^{2}} d \alpha^{2}+\frac{1}{1-\alpha^{2}} s_{A B} d \theta^{A} d \theta^{B}\,.
 \ee
  Here $\theta^{A}=(z,\bar{z})$ and $s_{AB}=2 P^{-2} d z d\bar z$.
 Note also that here $\alpha$ is related to $\tau$ used in \cref{eq:h-hyp-tau} by $\alpha:= \tanh \tau$, $-1 \leq \alpha \leq 1$ and $\alpha \to \pm 1$ corresponds to the limits to $\nulls^{\pm}$ .\\
 The vector field $\dd{X}^{a}$ can be written as

\be \label{eq:ddXa-exp}
    \dd X^a = Z \partial_\alpha + (P X) \partial_{\bar z} + (P \bar X) \partial_z\,,
\ee
where 
\be
    Z \wt s= 0 \eqsp X \wt s = -1\,.
\ee
As before, the spin weights are determined by the fact that $\dd{X}^{a}$ is invariant under spin transformations. Note also that \(Z\) is a real function while \(X\) is complex on \(\hyp\).
In these coordinates, the components of the equation \(\dd D_{(a}\dd X_{b)} =0 \) are given by
\begin{subequations}
\begin{align}
    0 &= (1-\alpha^2) \partial_\alpha Z + 2 \alpha Z \label{eq:first}\,, \\    
    0 &= (1-\alpha^2) \partial_\alpha X - \bar\eth Z \label{eq:second}\,,\\
    0 &= \bar\eth X \label{eq:third}\,, \\
    0 &= (1-\alpha^2) (\eth X + \bar{\eth}\, \bar X) + 2\alpha Z \label{eq:fourth}\,,
\end{align}
\end{subequations}
where the operators \(\eth\) and \(\bar\eth\) are defined in \cref{eq:ethethbaraction}. Note that \cref{eq:third} is the same equation that we encountered for $X$ in \cref{sec:bms-gen} which, in terms of spherical harmonics on the unit-sphere means that \(X\) is \(\ell = 1\) (or, stated in a conformally invariant way, that $\dd{X}^{A}$ is a conformal Killing vector field). Using this, \cref{eq:second} implies that \(Z\) is also \(\ell = 1\). As a result, the solution to \cref{eq:first} is given by
\be \label{eq:Z-soln}
    Z =  (1-\alpha^2) \sum_{m=-1}^{m=1} K_m Y_{\ell=1,m}^{s=0}\,,
\ee
where \(K_m\) labels three complex constants. Using \cref{eq:swsh-cc}, the fact that $Z$ is real relates these constants through 
\be \label{eq:Kconj}
    K_0 \in \bb R \eqsp K_{-1} = - \bar{K_1}\,,
\ee
leaving us with three independent real constants. Similarly, using
\be
    X = \sum_{m=-1}^{m=1} X_m(\alpha) Y_{\ell=1,m}^{s=-1} \,,
\ee
we see that \cref{eq:second} becomes
\be
    0 = \partial_\alpha X_m (\alpha) - K_m \,,
\ee
and so we have
\be
    X_m(\alpha) = R_m + \alpha K_m\,.
\ee
Moreover, \cref{eq:fourth} implies that
\be \label{eq:R-conj}
    R_m = (-)^{m+1} \bar {R_{-m}}\,.
\ee
As a result, $R_{m}$ also labels three real constants (with $m=-1,0,1$). We then see that \(K_m\) and \(R_m\) each represent three real numbers. They parametrize boosts and rotations respectively which can be seen by noting the fact that the divergence of $\dd{X}^{A}$ on a cross-section of $\hyp$, which is given by \(\eth X  + \bar\eth\, \bar X\),  is zero when $K_{m} =0\, \forall m$ and only non-zero when $\exists\, m: K_{m} \neq0$.

The expression for the Lorentz charge on $\hyp$ depends on $\dv X^{a}$ (see \cref{eq:AHcharge}), which is defined by $\dv X^a \defn \tfrac{1}{2} \dd\varepsilon^{abc} \dd D_b \dd X_c$. The properties of $\dv X^{a}$ were discussed in Appendix.~B of \cite{Prabhu:2019daz}. In particular, it was shown that it satisfies
\be \label{eq:dualXeq}
\dd{D}_{(a} \dv X_{b)} =0\,,
\ee
and therefore, $\dv X^{a}$ also represents a Lorentz symmetry on $\hyp$. 

To study the behavior of $\dv X^{a}$ in the limit to $\nulls^{\pm}$, we start with
\be \label{eq:star-op}
    \dv X^a  = ({}^\star Z) \partial_\alpha + P({}^\star X) \partial_{\bar z} + P(\bar{{}^\star X}) \partial_z\,.
\ee
Evaluating this explicitly and rewriting the resulting expression using \(\eth\) and \(\bar\eth\), we obtain  \footnote{In our conventions, the volume form on $\hyp$ in $(\alpha, z ,\bar{z})$ coordinates is given by $\dd{\varepsilon}_{abc} \equiv \frac{  2 i }{(1+z \bar{z})^{2} (1-\alpha^{2})^{2}} d\alpha \wedge dz \wedge d\bar{z}$.} 

\be\begin{aligned}
    {}^\star Z & = - \frac{i}{2} (1-\alpha^2) ( \eth X - \bar\eth \, \bar X )\,, \\
    {}^\star X & =  \frac{i}{2} \lb[ \bar{\eth} Z + (1-\alpha^2) \partial_\alpha X + 2\alpha X \rb] \\
    & =i \lb( \bar{\eth} Z + \alpha X \rb)\,,
\end{aligned}\ee
where the last equality uses \cref{eq:second}. Then, using the expressions for $Z$ and $X$ obtained above as well as \cref{eq:Kconj,eq:R-conj}, we obtain
\be \label{eq:dualveclimits}
    {}^\star Z =(1-\alpha^2) \sum_{m=-1}^{m=1} i R_m Y_{\ell=1,m}^{s=0} \eqsp {}^\star X = \sum_{m=-1}^{m=1} i (K_m + \alpha R_m ) Y_{\ell=1,m}^{s=-1}\,.
\ee
We therefore see that this ``dual'' transformation, $\dd{X}^{a} \to \dv{X}^{a}$, effectively interchanges \(K_m\) and \(R_m\) i.e. boosts and rotations. Note also that in the limit $\alpha \to \pm 1$, ${}^\star Z \to 0$ and one can check by explicitly evaluating the Hodge dual in $(z,\bar{z})$ coordinates that
\be \label{eq:dual-hdual-rel}
\dv X^{a}|_{\nulls^{\pm}} = -\dv{X}^{a}|_{\nulls^{\pm}}\,,
\ee
where $\dv{X}^{a}:= \tilde{\dd{\varepsilon}}_{b}{}^{a} \dd{X}^{b}$.

We now study the transformation of $\dd{X}^{a}$ under the reflection map $(\alpha, \theta^{A}) \to (-\alpha,-\theta^{A})$. Note that \(K_m\) and \(R_m\) are reflection-even since they are constants. From the transformation under the parity operation of the spin weighted spherical harmonics (\cref{eq:swsh-parity}), we can conclude that
\be
    Z(-\alpha,-\theta^A) = - Z(\alpha,\theta^A)\,,
\ee
as well as
\be\begin{aligned}
    \bar X(-\alpha, -\theta^A) & = \sum_{m=-1}^{m=1} \lb[ \bar{R_m} - \alpha \bar{K_m} \rb] \bar{ Y_{\ell=1,m}^{s=-1} }(-\theta^A) \\
    & = \sum_{m=-1}^{m=1} \lb[ (-)^{m+1} R_{-m} - \alpha (-)^m K_{-m} \rb] (-)^{m-1} Y_{\ell=1, -m}^{s=1}(-\theta^A)  \\
    & = e^{2 i\phi} \sum_{m=-1}^{m=1} \lb[ R_m + \alpha K_m \rb] Y_{\ell=1, m}^{s=-1}(\theta^A) = e^{2 i \phi} X(\alpha,\theta^A)\,,
\end{aligned}\ee
where $e^{2 i \phi} = z/\bar{z}$. Similarly, $X(-\alpha,-\theta^{A} )= e^{-2 i\phi} \bar{X}(\alpha,\theta^{A})$. Using the fact that under antipodal map $(\theta^{a} \to -\theta^{a}: z \to -1/\bar{z})$
\begin{align}
P \partial_{z} &\to  e^{-2 i\phi} P \partial_{\bar{z}}  \,,\notag\\
P \partial_{\bar{z}} &\to  e^{2 i\phi} P \partial_{z}\,,
\end{align}
we find that $\dd{X}^{a} \to \dd{X}^{a}$ under the reflection map and $\dd{X}^{a}$ is hence even under this map. It is straightforward to show that in the same way, $\dv X^{a}$ is also even under the reflection map. 

Recall from \cref{eq:Z-soln} that  \(Z\vert_{\nulls^\pm} = 0\) and therefore \(\dd X^a\) becomes tangent to \(\nulls^\pm\) in the limit.  Thus, on \(\nulls^\pm\) a spi symmetry is given by \((\dd F^\pm, \dd X^a)\) where \(\dd F = \dd\Sigma \dd f\) and $\dd{X}^{a}$ is a conformal Killing vector field tangent to $\nulls^{\pm}$.

\section{Fixing the supertranslation freedom at \(i^0\)}
\label{sec:fixing-st}

Our goal now is to show the matching of Lorentz charges at past and future null infinity in the limit to spatial infinity. We will show that this matching follows from requiring the continuity, at $\nulls^{\pm}$, of a quantity constructed from the Weyl tensor and a vector field in spacetime that limits to a BMS symmetry on $\scri$ with its BMS supertranslation part being zero and a spi symmetry on $\hyp$ with its spi supertranslation part being zero. This will define for us our notion of a ``pure'' Lorentz symmetry. Recall from the discussion of the asymptotic symmetry algebras in \cref{sec:bms-gen} and \cref{sec:spi-symm} that the Lorentz algebra forms  quotient subalgebras of $\mf{bms}$ and $\mf{spi}$ and therefore Lorentz symmetries are only defined as equivalence classes of symmetries that are related by supertranslations. Therefore, the notion of a ``pure'' Lorentz symmetry only makes sense when the supertranslation freedom is fixed. We will show below that restricting the conformal freedom in spatial directions near spatial infinity restricts the allowed spi supertranslations. This will then be used to restrict the allowed BMS supertranslations by requiring the continuity of a quantity constructed from $S_{ab}$ (defined in \cref{eq:S-defn}) in the limit to spatial infinity along both null and spatial directions. 

\paragraph*{The Ashtekar-Hansen gauge:}\label{rem:shear-fall-off}
 Recall from \cref{para:conf-choice} that we picked a conformal frame in a neighborhood of $i^{0}$ where $\Phi=2$. This choice fixes the dependence of $\omega$ on the null generators of $\scri$ near $i^0$. However, it does not restrict $\lim_{\to i^{0}} \alpha = \dd{\alpha}$ (defined below \cref{eq:conf-freedom}). One can use this freedom to do a conformal transformation such that $\dd{K}_{ab} \to \dd{K}_{ab} - 2 (\dd{D}_{a} \dd{D}_{b} \dd{\alpha} + \dd{h}_{ab} \dd{\alpha})=0$, as discussed in detail in Remark 6.3 of \cite{Prabhu:2019daz}. This is what we refer to as the ``Ashtekar-Hansen gauge'' since this choice was first made in \cite{AH}. Note that this does not exhaust the freedom in the choice of $\dd{\alpha}$. In particular, it leaves ``un-fixed'' the spi supertranslations  that satisfy $\dd{D}_{a} \dd{D}_{b} \dd{f} + \dd{h}_{ab} \dd{f} =0$ which precisely correspond to spi translations (see \cref{eq:translations-condition}). We will return to these at the end of this section.
 
  We now show that the condition $\dd{K}_{ab} =0$ can be used to restrict the supertranslation freedom at null infinity. Consider the quantity \(\Sigma^{-1} S_{ab} m^{a} m^{b}\) in a neighborhood of $i^0$ in the (unphysical) spacetime. We take its limits to $\nulls^{\pm}$ along both null and spatial directions and require that this quantity be continuous on $\nulls^{\pm}$. We take these limits along the null and spacelike hypersurfaces described in \cref{para:surfaces}. Here, $m^{a}$ along with its complex conjugate $\bar{m}^{a}$, forms an orthonormal basis on cross-sections of the surfaces described in \cref{para:surfaces} such that $m_a m^a = \bar{m}_{a} \bar{m}^{a} =0$, $m_{a} \bar{m}^{a} =1$ and $q_{ab}^{\prime}$, the metric on cross-sections, $S'$, of the surfaces described in \cref{para:surfaces}, is such that $q_{ab}^{\prime} = 2 m_{(a} \bar{m}_{b)}$. Note also that this metric limits to the intrinsic metric on cross-sections of $\scri$ and $\hyp$. On \(\scri\), we have
 \be \label{eq:sigma-Sab}
     \Sigma^{-1} S_{ab} m^a m^b = \Sigma S_{ab} \tilde m^a \tilde m^b\,,
 \ee
 where $\tilde{m}^{a}$ satisfies $\tilde{m}^{a} = \Sigma^{-1} m^{a}$ and is such that $\tilde{q}_{ab} = 2 \tilde{m}_{(a} \bar{\tilde{m}}_{b)}$. Recall that $\tilde{q}_{ab}$ is the rescaled metric that limits, along $\scri$ (\cref{eq:rescaled-q}), to a direction-dependent metric $\tilde{\dd{q}}_{ab}$ on $\nulls^{\pm}$. Next, note that in the limit to \(i^0\) along spatial directions, we have 
 \be \label{eq:rescaledK}
     \lim_{\to i^0} \Sigma^{-1} S_{ab} m^a m^b = \dd \Sigma^{-1} \dd S_{ab} \dd m^a \dd m^b =\dd \Sigma \dd K_{ab} \tilde{\dd m}^a \tilde{\dd m}^b = \dd{\Sigma} \STF \dd{K}_{ab} \,.
 \ee
  Here, the direction-dependent limit of $m^{a}$ to $i^{0}$ has been denoted by $\dd{m}^{a}$. Further, $\tilde{\dd{m}}^{a}$ is such that $\tilde{\dd{m}}^{a} = \dd{\Sigma}^{-1} \dd{m}^{a}$ and $\tilde{\dd{q}}_{ab}= 2 \tilde{\dd{m}}_{(a} \bar{\tilde{\dd{m}}}_{b)} $. Note also that the third equality in \cref{eq:rescaledK} follows from $\STF \dd{K}_{ab} = \STF \dd{S}_{ab}$ (which follows from \cref{eq:potentials-defn}).
We then see that when \(\dd K_{ab} = 0\), assuming continuity of $\Sigma^{-1} S_{ab} m^{a} m^{b}$ at $\nulls^\pm$ implies that in the limit to $\nulls^\pm$ along $\scri$, we have
 \be
     \lim_{\to \nulls^\pm} \Sigma S_{ab} \tilde m^a \tilde m^b = 0 \implies \lim_{\to \nulls^\pm} \STF (\Sigma S_{ab}) = 0 \,,
 \ee
 which, using
 \be
     N_{ab} \hateq \STF S_{ab} - 2 \Phi \sigma_{ab} \,,
 \ee
(which follows from \cref{eq:News-Ric} with $\tau_{a}\hateq0$) and the fact that $N_{ab} = O(\Sigma^{-(1+\epsilon)})$ as $\Sigma^{-1} \to 0$ along $\scri$ (see \cref{para:null-reg}),
 implies 
 \be \label{eq:shearfalloff}
     \lim_{\to \nulls^\pm} \Sigma\, \sigma_{ab} = 0 \,.
 \ee
Requiring this fall-off on $\sigma_{ab}$ reduces the supertranslation freedom at null infinity to translations since general supertranslations (that are not translations) do not preserve this fall-off.  

Although we have now fixed the supertranslation freedom at both null and spatial infinity, to unambiguously define a notion of ``pure'' Lorentz symmetry, we still have to contend with the translation freedom. To fix that, we proceed as follows. We consider a vector field, $\xi^{a}$, that limits to a BMS symmetry at $\scri$ and a spi symmetry on $\hyp$. We require that $\lim_{\to i^{0}} \Sigma \alpha_{(\xi)}= \lim_{\to i^{0}} \Omega^{-1} \Sigma\, \xi^{a}\, \nabla_{a} \Omega$ vanishes along both null and spatial directions. In the latter limit, this implies $\lim_{\to i^{0}} \Omega^{-1}\dd{\Sigma}\, \xi^{a} \eta_{a} =0$ which means $\dd{\Sigma} \dd{f} =0 \Rightarrow \dd{f} =0$ on $\hyp$. Therefore, this condition sets the translations that were left unfixed in going to the Ashtekar-Hansen gauge to zero and $\lim_{\to i^{0}} \Omega^{-\half} \xi^{a}$, subject to these conditions, gives us our notion of a ``pure'' Lorentz symmetry on $\hyp$. To see what $\lim_{\to i^{0}} \Sigma \alpha_{(\xi)}=0$ taken along $\scri$ implies, recall from \cref{eq:alpha-exp} that $\Sigma \alpha_{(\xi)} =\Sigma \Lie_{n} \beta= 2 \beta_{0}$. Since $\beta_{0}$ is independent of $\Sigma^{-1}$, requiring $\lim_{\to i^{0}} \Sigma \alpha_{(\xi)}=0$ along $\scri$ implies that it vanishes everywhere (in the neighborhood of  $i^{0}$ on $\scri$ where we have set $\Phi=2$). This gives us our notion of a ``pure'' Lorentz symmetry on $\scri$. Note that since we have only specified the asymptotic behavior of ``pure'' Lorentz symmetries, we are free to extend them into the spacetime in any way.  Denoting ``pure'' Lorentz symmetries on $\scri$ by $X^{a}$, we pick this extension to be one which satisfies  $\Sigma^{-2} K^{a} \nabla_{a} X^{b} \hateq 0$ where $K^{a}$ is defined by \cref{eq:startingpoint-1} and is the (affine) generator of null surfaces along which we will consider limits to $\scri$ (discussed in \cref{para:surfaces}). This will turn out to simplify some of our later calculations.

Having clarified what we mean by ``pure'' Lorentz, we will refer to these simply as Lorentz symmetries henceforth and to the associated charges as Lorentz charges. 
\begin{remark}[Matching of  Lorentz symmetries]\label{rem:matchingpurelorentz}
It follows from the analysis in \cref{sec:bms-gen,sec:LorentzsymmonH} that the Lorentz symmetries as defined above correspond to conformal killing vector fields on $\nulls^{\pm}$ in the limits along $\scri^{\pm}$ as well as $\cyl$. There is therefore an isomorphism between Lorentz symmetries at null and spatial infinity in this limit. Moreover, since, as shown in \cref{sec:LorentzsymmonH}, $\dd{X}^{a}$ is even under the reflection map on $\hyp$ which maps $\nulls^{-}$ to $\nulls^{+}$, we see that in the limit to spatial infinity, Lorentz symmetries at past and future null infinity match each other up to antipodal reflection.
\end{remark}
To recap, the fall-offs along $\scri$ that we will assume to hold are (for some small $\epsilon >0$)
 \begin{align} \label{eq:fall-offs}
  & N_{a b}=O\left(\Sigma^{-(1+\epsilon)}\right), \quad  \mathcal{R}_{a b}=O\left(\Sigma^{-(1+\epsilon)}\right)  \notag\\
  &\sigma_{ab} = O(\Sigma^{-(1+\epsilon)})\,,  \text { as } \Sigma^{-1}\to 0 \text { along } \scri\,.
  \end{align}
 where the first two were part of the definition of null-regular spacetimes (\cref{para:null-reg}) and the fall-off of $\sigma_{ab}$ follows from \cref{eq:shearfalloff}. These fall-offs imply certain conditions on the behavior of $\mc{P}^{\ast}$ in the limit $\Sigma^{-1} \to 0$ along $\scri$ which we turn to deriving next.
  \paragraph*{Fall-off of $\mc{P}^{\ast}$:}\label{rem:Pstar-falloff}
  In our foliation of $\scri$,  $\mc{P}^{\ast}$ (defined in \cref{eq:weyl-defn}) can be related to the shear and News tensors through \cite{WZ}
  \begin{equation} \label{eq:Pstar-identity}
    \mc P^* \hateq \varepsilon^{ab} \lb[ \ms D_a \ms D_c  \sigma_b{}^c - \tfrac{1}{2} N_{ac} \sigma_b{}^c \rb]\,.
  \end{equation}
  Let us consider its behavior as $\Sigma^{-1} \to 0$. Using \cref{eq:fall-offs}, we see that 
  \begin{subequations}\be
  \varepsilon^{ab}  \ms D_a \ms D_c  \sigma_b{}^c = O(\Sigma^{-(-3+\epsilon)})\,, \ee \be\varepsilon^{ab} N_{ac} \sigma_{b}{}^{c} = O(\Sigma^{-(-2+2\epsilon)})\,. \label{eq:term2falloff}
  \ee\end{subequations}
Note also that since $\sigma_{ab}$ is a symmetric and trace-free tensor, in terms of spherical harmonics on the unit-sphere, it is supported only on $\ell \geq 2$ tensor harmonics. Since $X^{b} q^{a}{}_{b}$ comprises purely $\ell=1$ vector spherical harmonics (as shown in \cref{sec:bms-gen}), as a consequence of the orthogonality of spherical harmonics, the first term in $\mc{P}^{\ast}$ drops out when integrated against $X^{b} q^{a}{}_{b}$ on a unit-sphere. Using this, we obtain that $\int_{S_{\Sigma}} \tilde{\varepsilon}_{2} \Sigma^{-2} X^{b} q^{a}{}_{b}\, \ms{D}_{a} \mc{P}^{\ast} =-\frac{1}{2}\int_{S_{\Sigma}} \tilde{\varepsilon}_{2} \Sigma^{-2} X^{b} q^{a}{}_{b}\, \varepsilon^{de}  \ms{D}_{a}( N_{dc} \sigma_{e}{}^{c}) $, where $\tilde{\varepsilon}_{2}$ is the unit area element and where we have implicitly used the fact that $S_{\Sigma}$ are $\Sigma=\text{constant}$ cross-sections and therefore factors of $\Sigma^{-1}$ can be pulled outside the integral over $S_{\Sigma}$. This goes to 0 as $\Sigma^{-1} \to 0$ because of \cref{eq:term2falloff} and the fact that $X^{b} q^{a}{}_{b}$ has a finite limit as $\Sigma^{-1} \to 0$, as shown in \cref{sec:bms-gen}. Hence, we have
  \be \label{eq:todiscardPstar}
  \lim_{\Sigma^{-1} \to 0}\int_{S_{\Sigma}} \tilde{\varepsilon}_{2} \Sigma^{-2} X^{b} q^{a}{}_{b}\, \ms{D}_{a} \mc{P}^{\ast} =0 \,.
  \ee
  
 \noindent The same argument also shows that

  \be \label{eq:todiscardPstar-2}
  \lim_{\Sigma^{-1} \to 0}\int_{S_{\Sigma}} \tilde{\varepsilon}_{2} \Sigma^{-2} X^{b} \varepsilon^{a}{}_{b}\, \ms{D}_{a} \mc{P}^{\ast} =0 \,.
  \ee
  One can contrast this with our assumed fall-off for $\mc{P}$ which can be deduced from \cref{eq:P-falloff} using \cref{eq:P-defn2} and is that the limit $ \lim_{\Sigma^{-1} \to 0}\Sigma^{-3} \mc{P}$ exists.
\section{Matching the Lorentz charges}\label{sec:matching}
 In this section, we consider limits of the Lorentz charges along $\scri^{\pm}$ and $\cyl$ to $\nulls^{\pm}$. A priori, the limits to $\nulls^{\pm}$ along $\scri^{\pm}$ and $\cyl$ are completely independent. However, we will show that there is a  tensorial quantity in spacetime that reduces to the limiting expressions for these charges in each of these limits. Therefore, if we assume that the aforementioned quantity is continuous at $\nulls^{\pm}$, the Lorentz charges at null and spatial infinity match in the limit to $\nulls^{\pm}$. We will then discuss how this leads to the matching of Lorentz charges between past and future null infinity. Comments on the validity of our continuity assumption are deferred to the next section.

 Consider first the charges at null infinity. In our chosen foliation, the expression for the charge associated with a BMS symmetry, $\xi^{a}$, on a finite cross-section, $S_{\Sigma}$, of $\scri^{+}$ is given by \cite{GPS} 
\be\label{eq:Q-GR-defn-2}
    \mc Q[\xi^a; S_{\Sigma}] \hateq - \frac{1}{8\pi} \int_{S_\Sigma} \df\varepsilon_2 \left[ \beta ( \mc P + \tfrac{1}{2}\sigma^{ab} N_{ab} ) + q^{a}{}_{b} X^b  \mc J_a + q^{a}{}_{b} X^b \sigma_{ab} \ms D_c \sigma^{bc} - \tfrac{1}{4} \sigma_{ab} \sigma^{ab} \ms D_c (q^{c}{}_{d} X^d) \right] \,,
\ee
where the Weyl tensor components appearing in the expression above were defined in \cref{eq:weyl-defn}, $\sigma_{ab}$ was defined in \cref{eq:sigma-defn}, $N_{ab}$ was defined in \cref{eq:News-defn} while $\beta$ and $X^{a}$ parametrize $\xi^{a}$ as in \cref{eq:BMSsymmonscri}. We now consider its behavior in the limit $\Sigma^{-1} \to 0$. First, solving the evolution equations for $\mc{P}$ and $\mc{J}_{a}$ given in  \cref{eq:weyl-evol} using $\tau_{a}\hateq 0$, $n^{a} \hateq 2 \Sigma^{-1} \partial_{\Sigma^{-1}}$, $\Phi=2$, the fall-offs given in \cref{eq:fall-offs} as well as \cref{eq:Pstar-identity,eq:weyl-News-2,eq:todiscardPstar-2,eq:rescaledareaelement}, we find that 
\be \label{eq:isfinite}
\lim_{\Sigma^{-1} \to 0} \int_{S_{\Sigma}} \tilde{\varepsilon}_{2} \Sigma^{-2} X^{b} q^{a}{}_{b} (\mc{J}_{a} + \frac{1}{4} \ms{D}_{a} \mc{P}) \, \mbox{is finite} \,,
\ee
where $\tilde\varepsilon_{2}$ denotes the unit area element and we have dropped terms that integrate to zero because of the orthogonality of spherical harmonics. We then take the limit $\Sigma^{-1} \to 0$ of \cref{eq:Q-GR-defn-2} using \cref{eq:fall-offs}. For a Lorentz symmetry (that is, where $\beta_{0} =0$), we see, using \cref{eq:beta-soln,eq:isfinite}, that in the limit to $\nulls^+$, the charge becomes 
\be\label{eq:Q-GR-on-N+}
    \mc Q[X^a; \nulls^+] \hateq - \frac{1}{8\pi} \int_{\nulls^+} \df\tilde{\varepsilon}_2 \Sigma^{-2} X^{b} q^{a}{}_{b} (\mc{J}_{a}+\frac{1}{4} \ms{D}_{a} \mc P) \,.
\ee
 It can be shown by similarly taking limits to $\nulls^-$ along $\scri^{-}$ that the charge associated with a Lorentz symmetry on $\nulls^-$ is given by
\be\label{eq:Q-GR-on-N-}
    \mc Q[X^a; \nulls^-] \hateq  -\frac{1}{8\pi} \int_{\nulls^-} \df\tilde{\varepsilon}_2 \Sigma^{-2} X^{b} q^{a}{}_{b} (\mc{J}_{a}+\frac{1}{4} \ms{D}_{a} \mc P) \,.
\ee
We now turn to the charges at spatial infinity. The charge associated with a Lorentz symmetry, $\dd{X}^{a}$, in Ashtekar-Hansen gauge on a cross-section ${\mc{S}}$ of $\hyp$, is given by \cite{AH,Prabhu:2019daz}
\be \label{eq:AHcharge}
\mc{Q}[\dd{X}^{a},\mc{S}] = -\frac{1}{8\pi}\int_{\mc{S}} \dd{\varepsilon}_{2} \dd{u}^{a} \dd{\beta}_{ab} \dv{X}^{b}\,,
\ee
where recall that $\dd{u}^{a}$ is the future-directed timelike normal on $\mc{S}$ and $\dv{X}^{b}$ was defined in \cref{eq:star-op}. It follows from \cref{eq:div-beta-2} and \cref{eq:dualXeq} that this charge is conserved on $\hyp$. This implies that $\dd{\beta}_{ab}$ has a definite behavior under the reflection map defined in \cref{eq:reflection-hyp}, which can be deduced as follows. Consider two cross-sections $\mc{S}_1$ and $\mc{S}_{2}$ of $\hyp$ such that $\Upsilon \circ \mc{S}_1 = \mc{S}_{2}$ where $\Upsilon \circ$ denotes the action of the reflection map. Then, charge conservation implies that
\begin{align}
0=\int_{\mc{S}_2} \dd\varepsilon_{2} \dd{u}^{a} \dd\beta_{ab} \dv{X}^{b} - \int_{\mc{S}_1} \dd\varepsilon_{2} \dd{u}^{a} \dd\beta_{ab} \dv{X}^{b} &= \int_{\mc{S}_2}  \big[ \dd\varepsilon_{2} \dd{u}^{a} \dd\beta_{ab} \dv{X}^{b}- \Upsilon \circ (\dd{\varepsilon}_{2} \dd{u}^{a} \dd{\beta}_{ab} \dv{X}^{b})\big]\,, \notag \\
&=\int_{\mc{S}_2}  \big[ \dd\varepsilon_{2} \dd{u}^{a} \dd\beta_{ab} \dv{X}^{b}+ \dd{\varepsilon}_{2} \dd{u}^{a} \Upsilon \circ (\dd{\beta}_{ab} \dv{X}^{b})\big]\,,
\end{align}
where in the last equality, we have used the fact that $\Upsilon \circ (\dd{\varepsilon}_{2} \dd{u}^{a} ) = - \dd{\varepsilon}_{2}  \dd{u}^{a} $. This follows because $\Upsilon \circ \dd{\varepsilon}_{2} =  - \dd{\varepsilon}_{2}$ and because $\dd{u}^{a}$ is future-directed on both $\mc{S}_1$ and $\mc{S}_2$ and therefore does not get acted on by the reflection map. Moreover, since $\dv{X}^{a}$ is even under the reflection map on $\hyp$ (as shown in \cref{sec:LorentzsymmonH}), it follows that the charge only receives contribution from reflection-odd solutions of $\dd{\beta}_{ab}$. Using this as motivation, in the rest of this paper we will restrict our attention to only reflection-odd solutions for $\dd{\beta}_{ab}$ (derived in \cref{sec:beta-soln}).\footnote{This condition on $\dd{\beta}_{ab}$ is satisfied in the Kerr-Newman family of spacetimes; see, e.g, Appendix. C of \cite{AH}.} Since we have specialized to spacetimes where $\dd{E}_{ab}$ is reflection-even as remarked in \cref{para:null-reg}, it follows from \cref{eq:wave-eqn-beta-2} and the fact that the reflection map preserves the volume form on $\hyp$ that the reflection-odd solutions for $\dd{\beta}_{ab}$ satisfy the equation
\be \label{eq:waveeqn-beta}
(\dd{D}^{2} - 2) \dd{\beta}_{ab}=0\,.
\ee
 In fact, one can show by solving for the reflection-even solutions to this equation in the same way as for the reflection-odd solutions in \cref{sec:beta-soln} that the $\ell=1$ reflection-even solution diverges in the limits to $\nulls^{\pm}$ and therefore for these solutions the Lorentz charge diverges in these limits as well. While we have not shown it explicitly, we expect this property to remain true even for the reflection-even solutions to \cref{eq:wave-eqn-beta-2}. Since these solutions are clearly pathological, this serves as further motivation for discarding the reflection-even solutions for $\dd{\beta}_{ab}$.

 As shown in \cref{sec:beta-soln} and \cref{sec:spi-symm}, reflection-odd solutions for  $\dd{\beta}_{ab}$ and $\dv X^{a}$ have a finite limit to $\nulls^{\pm}$. Using this and  \cref{eq:nulls-area,eq:relbetnormals}, we see that the limit of the charge in \cref{eq:AHcharge} to $\nulls^{\pm}$ is non-vanishing and is given by

\be \label{eq:AHchargeonnulls}
\mc{Q}[\dd{X}^{a},\nulls^{\pm}] = - \frac{1}{8\pi}\int_{\nulls^{\pm}}  \tilde{\dd{\varepsilon}}_{2} \dd{\Sigma}^{-1} \dd{U}^{a} \dd{\beta}_{ab} \dv{X}^{b}=  \frac{1}{8\pi}\int_{\nulls^{\pm}}  \tilde{\dd{\varepsilon}}_{2} \dd{\Sigma}^{-1} \dd{U}^{a} \dd{\beta}_{ab}  \tilde{\dd{\varepsilon}}_{c}{}^{b}\dd{X}^{c}
\ee
where the last equality uses \cref{eq:dual-hdual-rel}. 

  Let us now consider the following quantity in a neighborhood of $i^{0}$ in the unphysical spacetime,
\be \label{eq:beta-component}
    -\frac{1}{8\pi} \int_{\mc{S'}}\, \tilde{\varepsilon}_{2}\, \Omega^{-\half} \Sigma^{-1}\,\ast C_{acbd}\,U^{a} (\nabla^{c} \Omega^{\half}) (\nabla^{d} \Omega^{\half})  (\Omega^{-\half} X^{b})\,,
\ee
where, as before, $\tilde{\varepsilon}_{2}$ is the unit area element, $L^{a}$ was defined in \cref{eq:Lamodified} and $U^{a}$ is a $C^{>-1}$ vector field at $i^{0}$, is defined as 
   \be \label{eq:Ua}
    U^{a} = L^{a}- N^{a} (-\Omega^{-1} \Sigma^{-2} + \frac{1}{2} \nabla_{b} \Sigma^{-1} \nabla^{b} \Sigma^{-1})\,,
     \ee 
  whose limit to $\hyp$ satisfies $\dd{U}^{a}=\dd{h}^{a}{}_{b} \dd{L}^{b}$\,. Further, $X^{a}$ is such that $\Omega^{-\half} X^{a}$ limits to a Lorentz symmetry on $\hyp$ and $X^{a}$ limits to a Lorentz symmetry on $\scri$. We now explore how this quantity behaves in the limit to $\nulls^\pm$ along $\cyl$ and along $\scri^\pm$. 

\subsection{Limit to $\nulls^{\pm}$ along $\cyl$}
  Consider first the limit of \cref{eq:beta-component} to cross-sections of $\hyp$ along the sequence of spacelike hypersurfaces described in \cref{para:surfaces}. Using $\lim_{\to i^{0}} \Omega^{-\half} X^{a} = \dd{X}^{a}$ in addition to \cref{eq:eta-defn} and \cref{eq:dd-Sigma}, we see that in the limit to $\hyp$, \cref{eq:beta-component} becomes 
    \be
  -\frac{1}{8\pi}  \int_{\mc{S}} \tilde{\dd{\varepsilon}}_{2} \dd{\Sigma}^{-1} \ast \dd{C}_{acbd} \dd\eta^{c} \dd\eta^{d} \dd U^{a} \dd{X}^{b}\,,
    \ee
 where $\mc{S}$ is a cross-section of $\hyp$. Using \cref{eq:beta-defn}, we obtain  
  \be \label{chargelimitspatial}
 \lim_{\to \nulls^{\pm}}  \frac{-1}{8\pi}\int_{\mc{S}} \tilde{\dd{\varepsilon}}_{2}\,\dd{\Sigma}^{-1} \ast \dd{C}_{acbd} \dd\eta^{c} \dd\eta^{d} \dd U^{a} \dd{X}^{b} =-\frac{1}{8\pi} \int_{\nulls^{\pm}} \tilde{\dd{\varepsilon}}_{2} \dd{\Sigma}^{-1} \dd{U}^{a} \dd{\beta}_{ab} \dd{X}^{b}\,,
 \ee  
This, from \cref{eq:AHchargeonnulls}, corresponds to the Lorentz charge, on $\nulls^{\pm}$, for any Lorentz symmetry given by $\tilde{\dd{\varepsilon}}_{b}{}^{a}\dd{X}^{b}$.

\subsection{Limit to $\nulls^{\pm}$ along $\scri^{\pm}$} 

Consider now the limit of \cref{eq:beta-component} to $\nulls^+$ along $\scrp$.\footnote{The reader who wishes to skip the details of this somewhat involved calculation may jump directly to \cref{eq:finaleq} where the final expression is given.} Throughout this calculation, we will often implicitly use \cref{eq:weyl-defn}, the symmetries of the Weyl tensor and the fact that under the replacement $C_{abcd} \to \ast C_{abcd}$, the expressions for the Weyl tensor components in \cref{eq:weyl-defn} change as follows
\be \label{eq:changes}
    \mc{P} \to \mc{P}^{\ast}\,, \quad  \mc{S}_{a} \to \mc{S}_{b}\, \varepsilon^{b}{}_{a}\,, \quad  \mc{J}_{a} \to \mc{J}_{b}\, \varepsilon_{a}{}^{b}\,, \quad \mc{R}_{ab} \to \mc{R}_{a}{}^{c}\, \varepsilon_{cb}\,.
\ee
As described in \cref{para:surfaces}, we will take the limit to cross-sections $S_{\Sigma}$ of $\scrp$ along a sequence of null hypersurfaces, that exist in a neighborhood of $i^{0}$, that intersect these cross-sections. We will then take the limit $\Sigma^{-1} \to 0$ along $\scrp$ which will define for us the limit of our expression to $\nulls^+$. We define
\be \label{eq:startingpoint-1}
    K^{a} \defn l^{a} - \Omega \alpha^{a} \,,
\ee  where $K^{a}$ is the affine null generator of the aforementioned null surfaces. Here, $\alpha^{a} \hateq -l^{b} \nabla_{b} l^{a}$. As indicated, this expression only fixes $\alpha^{a}$ at $\scrp$ and its expression away from $\scrp$ is chosen to ensure that $K^{a} \nabla_{a} K^{b} =K^{a} K_{a}=0$ all along the null surfaces. Converting the integrand in \cref{eq:beta-component} into quantities defined on $\scri^{+}$ using \cref{eq:N-defn,eq:Ua,eq:l-defn} (and relabeling the indices for later convenience), we get
    \begin{align}
   \Omega^{-\half} \Sigma^{-1}\,\ast C_{acbd}\,U^{a} (\nabla^{c} \Omega^{\half}) (\nabla^{d} \Omega^{\half})  (\Omega^{-\half} X^{b})=  \frac{\Omega^{-2} \Sigma^{-2}}{2} \ast C_{bcde}\,n^{d}\,n^{b}\, l^{c}\,X^{e}\,.
    \end{align}
       Using $\alpha^{a} \hateq -l^{b} \nabla_{b} l^{a}$, it follows that
       \be \label{eq:Kdotn}
       K^{a}\,n_{a}=-1 + O(\Omega^{2})\,.
       \ee
        We can then write
   \begin{align}\label{maineq-3-0}
   & \lim_{S^{\prime} \to S_{\Sigma}}\,\frac{1}{2} \int_{S^{\prime}}\, \tilde\varepsilon_{2}\, \Omega^{-2}\,\Sigma^{-2}\,\ast C_{bcde}\, n^{d}\,n^{b} l^{c}\, X^{e}= \lim_{S^{\prime} \to S_{\Sigma}}\,\frac{1}{2} \int_{S^{\prime}}\,\tilde\varepsilon_{2}\, \Omega\, K^{a} \nabla_{a} (\Sigma^{-2}\,\Omega^{-2} \ast C_{bcde}\, n^{d}\,n^{b}\, l^{c}\, X^{e})\notag\\
   & -\lim_{S^{\prime} \to S_{\Sigma}}\,\frac{1}{2} \int_{S^{\prime}} \tilde\varepsilon_{2}\, K^{a} \nabla_{a} (\Omega^{-1}\,\Sigma^{-2}\, \ast C_{bcde}\,n^{d}\,n^{b}\,l^{c}\,X^{e})\,.
   \end{align}
 Consider the first term on the right hand side of \cref{maineq-3-0}. This can be written as
   \begin{align} \label{eq:Winicour}
 &\lim_{S^{\prime} \to S_{\Sigma}}\,\frac{1}{2} \int_{S^{\prime}} \tilde \varepsilon_{2}\, \Omega\, K^{a} \nabla_{a} (\Sigma^{-2}\,\Omega^{-2} \ast C_{bcde}\, n^{d}\,n^{b}\, l^{c}\, X^{e})\notag \\
 &=\lim_{S^{\prime} \to S_{\Sigma}}\, \frac{\Omega}{2}\int_{ S^{\prime} } \tilde \varepsilon_{2}\, K^{a} \nabla_{a} (\Sigma^{-2}\,\Omega^{-2} \ast C_{bcde}\, n^{d}\,n^{b}\, l^{c}\, X^{e})\notag\\
 &=  \lim_{\Omega \to 0} \Omega \frac{d}{d\Omega}\big(\int_{S^{\prime}} \frac{\tilde \varepsilon_{2}}{2} \Sigma^{-2} \Omega^{-1} \mc{S}_{a}\, \varepsilon^{a}{}_{b}\, X^{b}\big)\,
 -\lim_{S^{\prime}\to S_{\Sigma}}\,\frac{1}{2}  \int_{S^{\prime}} \tilde\varepsilon_{2}\, \Sigma^{-2}\, \vartheta(K^{a})\, \mc{S}_{c}\, \varepsilon^{c}{}_{b} X^{b}\,,
    \end{align}
where in the first equality, we used the fact that $S^{\prime}$ denote $\Omega=\text{constant}$ cross-sections to move $\Omega$ outside the integral and in the second and third equalities, we used \cref{eq:changes} and Eq.~2.23 of \cite{Winicour-some-total-inv}, which, translated into our notation, states that\footnote{Note that the definition of expansion used in this paper (\cref{eq:expansion-defn}) is twice the definition in Eq.~2.25 of \cite{Winicour-some-total-inv}.}
\be
 \frac{d}{d\Omega} \int_{S^{\prime}}\tilde{\epsilon}_{2}\, B = \int_{S^{\prime}} \tilde{\epsilon}_{2}\, \big[(K^{a} \nabla_{a} B + \vartheta(K^{a}) B) (-n_{b} K^{b})^{-1}\big]\,,
\ee
for some scalar $B$ that has a finite integral over cross-sections $S^{\prime}$ as $S^{\prime} \to S_{\Sigma}$. Additionally, we have used \cref{eq:Kdotn}, along with the fact that the limit $S' \to S_{\Sigma}$ coincides with $\Omega \to 0$. Note that the expression in the round brackets in the first term in \cref{eq:Winicour} is finite on $\scrp$ for the following reason. Since $\mc{S}_{a} \hateq \frac{1}{2} \ms{D}^{b} N_{ab}$ (\cref{eq:weyl-News-2}), using \cref{eq:divsigma=0}, up to a total derivative term that would drop out upon integrating over $S_{\Sigma}$, $ \tilde\varepsilon_{2} \Sigma^{-2} \mc{S}_{a}\, \varepsilon^{a}{}_{b}\, X^{b}  =-\tfrac{\tilde  \varepsilon_{2}\,\Sigma^{-2}}{2} N_{ab}\, \ms{D}^{a} \varepsilon^{b}{}_{c} X^{c}$.  Since $\varepsilon^{a}{}_{b} X^{b}$ is a conformal killing vector on $S_{\Sigma}$ (as shown \cref{sec:bms-gen}), this vanishes upon contraction with $N_{ab}$ since $N_{ab}$ is a symmetric traceless tensor. As a result, $\int_{S^{\prime}} \tilde\varepsilon_{2} \Sigma^{-2} \mc{S}_{a}\, \varepsilon^{a}{}_{b}\, X^{b}$ is $O(\Omega)$ and hence 
    \be
   \lim_{\Omega \to 0} \Omega \frac{d}{d\Omega}\big(\int_{S^{\prime}} \frac{\tilde \varepsilon_{2}}{2} \Sigma^{-2} \Omega^{-1} \mc{S}_{a}\, \varepsilon^{a}{}_{b}\, X^{b}\big)=0\,.
   \ee
  As a result, we have
    \begin{align}\label{maineq-3}
  &\lim_{S^{\prime} \to S_{\Sigma}}\,\frac{1}{2} \int_{S^{\prime}}\, \tilde \varepsilon_{2}\, \Omega^{-2}\,\Sigma^{-2}\,\ast C_{bcde}\, n^{d}\,n^{b} l^{c}\, X^{e}   =\lim_{S^{\prime}\to S_{\Sigma}}\,\frac{-1}{2}\int_{S^{\prime}} \tilde \varepsilon_{2}\, \Sigma^{-2}\, \vartheta(K^{a})\, \mc{S}_{c}\, \varepsilon^{c}{}_{b} X^{b}\notag\\
     & - \lim_{S^{\prime} \to S_{\Sigma}}\, \frac{1}{2} \int_{ S^{\prime}} \tilde \varepsilon_{2}\, K^{a} \nabla_{a} (\Omega^{-1}\,\Sigma^{-2}\, \ast C_{bcde}\,n^{d}\,n^{b}\,l^{c}\,X^{e})\,.
     \end{align} 
      Using $\Sigma^{-2} K^{a} \nabla_{a} X^{b} \hateq 0$ (recall the discussion at the end of \cref{sec:fixing-st}), the second term on the right hand side above becomes
 \be\label{eq:exp-above}
  \lim_{S^{\prime} \to S_{\Sigma}}\,\frac{-1}{2} \int_{S^{\prime}} \tilde \varepsilon_{2}\,X^{e}\, K^{a} \nabla_{a} (\Omega^{-1}\,\Sigma^{-2}\, \ast C_{bcde}\,n^{d}\,n^{b}\,l^{c})\,.
 \ee
 Since $X^{e}$ is tangent to $\scrp$, we write it as $X^{e} =-n^{e} l_{b} X^{b} + q^{e}{}_{b} X^{b} + O(\Omega) $, where
 \be \label{eq:q}
  q^{e}{}_{b} = \delta^{e}{}_{b} + n^{e} l_{b} + n_{b} l^{e}\,.
  \ee
 Then \cref{eq:exp-above} becomes
 \begin{align}
 & \lim_{S^{\prime} \to S_{\Sigma}}\,\frac{-1}{2} \int_{S^{\prime}} \tilde \varepsilon_{2}\,(-n^{e} l_{f} X^{f} + q^{e}{}_{f} X^{f})\, K^{a} \nabla_{a} (\Omega^{-1}\,\Sigma^{-2}\, \ast C_{bcde}\,n^{d}\,n^{b}\,l^{c})\notag\\
 &= \lim_{S^{\prime} \to S_{\Sigma}}\,\frac{-1}{2} \int_{S^{\prime}} \tilde \varepsilon_{2}\,\,\Omega^{-1}\,\Sigma^{-2}\, \ast C_{bcde}\,n^{d}\,n^{b}\,l^{c}\,\big[{l_{f} X^{f} K^{a} \nabla_{a} n^{e}}- X^{f}\,K^{a} \nabla_{a}  q^{e}{}_{f} \big] \notag \\
 & -\lim_{S^{\prime} \to S_{\Sigma}}\,\frac{1}{2} \int_{S^{\prime}} \tilde \varepsilon_{2}\,  X^{f}\, K^{a} \nabla_{a}(\Omega^{-1}\,\Sigma^{-2} \ast C_{bcde}\,n^{d}\,n^{b}\,l^{c}\,q^{e}{}_{f})\,.
 \end{align}
 Using $\nabla_{a} n_{b} \hateq 2 g_{ab}$ (from \cref{eq:n-Phi}, adapted to $\Phi=2$) and \cref{eq:q}, the right hand side can be expanded out and written as
  \begin{align}\label{eq:piece-0}
 \lim_{S^{\prime} \to S_{\Sigma}} \int_{S^{\prime}} \tilde \varepsilon_{2}\,\,\Omega^{-1}\,\Sigma^{-2}\, \ast C_{bcde}\,n^{d}\,n^{b}\,l^{c}\, l^{e}\,  K_{f} X^{f} -\lim_{S^{\prime} \to S_{\Sigma}}\,\frac{1}{2} \int_{S^{\prime}} \tilde \varepsilon_{2}\, X^{f}\, K^{a} \nabla_{a}(\Omega^{-1}\,\Sigma^{-2}\,  \ast C_{bcde}\,n^{d}\,n^{b}\,l^{c}\,q^{e}{}_{f})\,,
 \end{align}
 where in simplifying this expression, we used $X^{a} n_{a} \hateq 0$.
  Note that since $K^{a} \hateq l^{a}$ and $l_{a} X^{a} \hateq \frac{1}{4} \ms{D}_{a} (q^{a}{}_{b} X^{b})$ (from \cref{eq:BMSsymmonscri,eq:beta-soln}), the first term above becomes 
  \be \label{eq:piece-1}
  \frac{1}{4} \int_{ S_{\Sigma}}\tilde \varepsilon_{2}\,\,\Sigma^{-2}\,\mc{P}^{\ast} \ms{D}_{a} (q^{a}{}_{b} X^{b}) = -
   \frac{1}{4} \int_{ S_{\Sigma}}\tilde \varepsilon_{2}\,\,\Sigma^{-2}\, q^{a}{}_{b} X^{b} \,\ms{D}_{a}\mc{P}^{\ast}\,,
   \ee
  where, because the integrand of this term is finite on $\scrp$, we took the limit and evaluated it directly on $S_{\Sigma}$ and  in the last step we did an integration by parts. Note that as we take $\Sigma^{-1} \to 0$ on $\scrp$, this term will go to zero from \cref{eq:todiscardPstar}. As a result, we can discard this term. Turn now to the second term in \cref{eq:piece-0}
  \begin{align} \label{eq:intermed-eq}
  \lim_{S^{\prime} \to S_{\Sigma}}\,\frac{-1}{2} \int_{S^{\prime}} \tilde \varepsilon_{2}\, X^{f}\, K^{a} \nabla_{a}( \Omega^{-1}\,\Sigma^{-2}  &\ast C_{bcde}\,n^{d}\,n^{b}\,l^{c}\,q^{e}{}_{f}) \notag \\
  & = \lim_{S^{\prime} \to S_{\Sigma}}\,\frac{-1}{2} \int_{S^{\prime}} \tilde \varepsilon_{2}\, X^{f}\, K^{a} \nabla_{a}(\Omega^{-1}\,\Sigma^{-2} \ast C_{bcde}\, l^{d} n^{c} n^{e} q^{b}{}_{f})\,,
  \end{align}
where we have simply relabeled the indices for later convenience. To evaluate this term, consider the Bianchi identity for the Hodge dual of the Weyl tensor.
  \be 
  \nabla_{[a} (\Omega^{-1} \ast C_{bc]de}) = 0\,,
  \ee
which can be rewritten as
   \begin{align} \label{eq:bianchi-calc-2}
   &\nabla_{a} (\Omega^{-1}\,\Sigma^{-2}\, \ast C_{bcde}) + \nabla_{c} (\Omega^{-1}\,\Sigma^{-2}\, \ast C_{abde}) + \nabla_{b} (\Omega^{-1}\,\Sigma^{-2}\,\ast C_{cade})-\Omega^{-1} \ast C_{bcde} \nabla_{a} \Sigma^{-2} \notag \\
   &-\Omega^{-1} \ast C_{cade} \nabla_{b} \Sigma^{-2} - \Omega^{-1} \ast C_{abde} \nabla_{c} \Sigma^{-2}=0\,.
   \end{align}
  Contracting this equation with $K^{a}\,l^{d}\,n^{c}\,n^{e}\,q^{b}{}_{f}$, we obtain
   \begin{align} \label{maineq2-2}
   &\frac{1}{2}K^{a} \nabla_{a} (\Omega^{-1}\Sigma^{-2} \ast C_{bcde}\, l^{d} n^{c} n^{e} q^{b}{}_{f}) = \frac{1}{2}\bigg[K^{a} \Omega^{-1}\,\Sigma^{-2}\, \ast C_{bcde} \nabla_{a} (l^{d}n^{c}n^{e}q^{b}{}_{f})\notag \\
   &-n^{c} \nabla_{c} (\Omega^{-1}\Sigma^{-2} \ast C_{abde}\, K^{a}l^{d}n^{e}q^{b}{}_{f})
   - q^{b}{}_{f} \nabla_{b}(\Omega^{-1}\Sigma^{-2} \ast C_{cade}K^{a}l^{d}n^{c}n^{e}) \notag\\&+ \Omega^{-1}\Sigma^{-2} \ast C_{abde} n^{c}\nabla_{c} (K^{a}l^{d}n^{e}q^{b}{}_{f}) + q^{b}{}_{f}\, \Omega^{-1}\,\Sigma^{-2}\, \ast C_{cade} \nabla_{b} (K^{a}l^{d}n^{c}n^{e})\notag\\
   & + \Omega^{-1} \ast C_{bcde} K^{a}l^{d}n^{c}n^{e}q^{b}{}_{f} \nabla_{a}\,\Sigma^{-2} + \Omega^{-1} \ast C_{cade}K^{a}l^{d}n^{c}n^{e}q^{b}{}_{f} \nabla_{b} \Sigma^{-2}+ \notag \\
   &\Omega^{-1} \ast C_{abde} K^{a} l^{d} n^{e} q^{b}{}_{f} n^{c}\nabla_{c} \Sigma^{-2} \bigg]\,.
   \end{align}
  Each of the terms on the right hand side of this expression is individually finite on $\scrp$ and therefore we can evaluate this expression directly on $\scrp$.  Using $n^{c} \nabla_{c}\, \varepsilon_{a}{}^{b}  \hateq 0$ (which holds since $\tau_{a} \hateq 0$), \cref{eq:weyl-defn}, \cref{eq:changes}, $K^{a} \hateq l^{a}$ (from \cref{eq:startingpoint-1}) and  $\Sigma \pounds_{n} \Sigma^{-1} \hateq 2$ (\cref{cond:Sigma-choice}), the two terms on the second line in this expression combine with the second term on the fourth line to give 
    \be \label{eq:inteq-1}
    -\frac{\varepsilon_{b}{}^{a}}{2} n^{c} \nabla_{c}( \Sigma^{-2}\,\mc{J}_{a}) + \frac{\Sigma^{-2}}{2}  \ms{D}_{b} \mc{P}^{\ast}=  \frac{\varepsilon_{b}{}^{a}}{2}(2\Sigma^{-2} \mc{J}_{a} -\Sigma^{-2} \pounds_{n} \mc{J}_{a}- 4  \Sigma^{-2} \mc{J}_{a})+ \frac{\Sigma^{-2}}{2} \,\ms{D}_{b} \mc{P}^{\ast}\,.
    \ee
        This can be simplified using the evolution equation for $\mc{J}_{a}$  (\cref{eq:evol-J} with $\tau_{a} \hateq 0$) and we obtain
   \begin{align}
  & -\frac{\varepsilon_{b}{}^{a}}{2}  n^{c} \nabla_{c}( \Sigma^{-2}\,\mc{J}_{a}) + \frac{\Sigma^{-2}}{2}   \ms{D}_{b} \mc{P}^{\ast} \notag\\
   &=  \Sigma^{-2}\,\varepsilon_{b}{}^{a} \mc{J}_{a}+ \frac{\Sigma^{-2}}{4}\,\ms{D}_{b} \mc{P}^{\ast} -\frac{\Sigma^{-2} \varepsilon_{b}{}^{a}}{4} \ms{D}_{a}\mc{P} + \Sigma^{-2}\,\varepsilon_{b}{}^{c} \sigma_{c}{}^{a} \mc{S}_{a}\,.
   \end{align}
    We now evaluate the remaining terms in \cref{maineq2-2}. In what follows, we drop terms proportional to $\Sigma^{-2} \ms{D}_{a} \mc{P}^{\ast}$ everywhere. This is because they contribute terms proportional to $\int_{S_{\Sigma}} \tilde\varepsilon_{2}\, \Sigma^{-2}\, X^{a} q_{a}{}^{b} \ms{D}_{b} \mc{P} ^{\ast}$ to \cref{eq:intermed-eq} and therefore, from \cref{eq:todiscardPstar}, drop out in the limit $\Sigma^{-1} \to 0$ which we take in the end. We use $\cdots$ to indicate that these terms have been suppressed in our expressions. We obtain, altogether, that
   \begin{align} 
   &\frac{1}{2} K^{a} \nabla_{a} (\Omega^{-1}\Sigma^{-2}\ast C_{bcde}l^{d} n^{c} n^{e} q^{b}{}_{f}) \hateq  -\Sigma^{-2} \varepsilon_{f}{}^{a} \mc{J}_{a}-\frac{\Sigma^{-2} }{4} \varepsilon_{f}{}^{a} \ms{D}_{a}\mc{P}+
      \Sigma^{-2}\,\varepsilon_{f}{}^{a}\sigma_{a}{}^{b} \mc{S}_{b} \notag \\
      & - \frac{\Omega^{-1}}{2}\,\Sigma^{-2} \ast C_{bcde}\, q^{b}{}_{f} \,n^{c}\,n^{e}\,\alpha^{d}+\Sigma^{-2} \mc{S}_{a}\, \varepsilon_{b}{}^{a}\, \sigma^{b}_{f} - \frac{\Sigma^{-2}}{2}\, \vartheta(l^{a})\varepsilon^{b}{}_{f}\, \mc{S}_{b}\, \notag\\
   &  + \frac{\Omega^{-1}}{2}\ast C_{bcde} l^{d} n^{c} n^{e} q^{b}{}_{f} K^{a} \nabla_{a} \Sigma^{-2}+\cdots\,.
   \end{align}
    Note that since $\varepsilon_{ab}$ is antisymmetric and $\sigma_{ab}$ is symmetric and trace-free, $\varepsilon_{ba} \sigma^{a}{}_{c}$  is a symmetric tensor (see, e.g, Appendix.~D of \cite{GPS} for a proof). Using this, the third term in the first line above cancels with the second term in the second line. Next, using $K^{a} \nabla_{a} \Sigma^{-2} \hateq l^{a} \nabla_{a} \Sigma^{-2}\hateq -\frac{1}{2}\nabla_{a} \Sigma^{-1} \nabla^{a} \Sigma^{-1}$ (where the last equality follows from \cref{eq:l-defn,eq:Lamodified,eq:N-defn,cond:Sigma-choice}), along with \cref{eq:weyl-defn,eq:changes} in the last term, we get
\begin{align}\label{eq:final2-2}
   &\frac{1}{2} K^{a} \nabla_{a} (\Omega^{-1}\Sigma^{-2}\ast C_{bcde}l^{d} n^{c} n^{e} q^{b}{}_{f}) \hateq - \Sigma^{-2}\,\varepsilon_{f}{}^{a} \mc{J}_{a}-\frac{\Sigma^{-2}}{4}  \varepsilon_{f}{}^{a}\, \ms{D}_{a} \mc{P} \notag \\  
    &- \frac{\Omega^{-1}}{2}\,\Sigma^{-2} \ast C_{bcde}\, q^{b}{}_{f} \,n^{c}\,n^{e}\,\alpha^{d}- \frac{\Sigma^{-2}}{2}\, \vartheta(l^{a})\,\varepsilon^{b}{}_{f}\,\mc{S}_{b}
     -\frac{\varepsilon^{b}{}_{f}}{4} \,S_{b} \nabla_{a} \Sigma^{-1} \nabla^{a} \Sigma^{-1}+\cdots\,.
   \end{align}  
    Let us now consider the term $- \frac{\Omega^{-1}}{2}\,\Sigma^{-2} \ast C_{bcde}\, q^{b}{}_{f} \,n^{c}\,n^{e}\,\alpha^{d}$ in \cref{eq:final2-2}. Using $\alpha^{a} \hateq -l^{b} \nabla_{b} l^{a}$ as well as \cref{eq:weyl-defn,eq:changes}, this term can be written as
    \be \label{eq:junk-term-2}
    -\frac{\Sigma^{-2}\, \varepsilon^{c}{}_{f}}{2} (n_{b}\, \mc{S}_{c} l^{a} \nabla_{a} l^{b} - \mc{R}_{bc}\, l^{a} \nabla_{a} l^{b})\,.
    \ee
 Then using \cref{eq:q-proj-affinity,eq:affin-phi2} to simplify this, we obtain              
\be
    \frac{1}{2} K^{a} \nabla_{a} (\Omega^{-1} \Sigma^{-2} \ast C_{bcde} l^{d} n^{c} n^{e} q^{b}{}_{f}) \hateq - \Sigma^{-2} \varepsilon_{f}{}^{a} \mc{J}_{a} -\frac{\Sigma^{-2} }{4} \varepsilon_{f}{}^{a}\,\ms{D}_{a} \mc{P} - \frac{\Sigma^{-2}}{2}\vartheta(l^{a})\,  \varepsilon^{b}{}_{f}\,\mc{S}_{b}  + \cdots\,.
\ee
Putting all of this together, we have                             
\begin{align}\label{maineq-3-2}
&\lim_{S^{\prime} \to S_{\Sigma}}\,\frac{1}{2} \int_{S^{\prime}}\, \tilde \varepsilon_{2}\, \Omega^{-2}\,\Sigma^{-2}\,\ast C_{bcde}\, n^{d}\,n^{b} l^{c}\, X^{e}   =\lim_{S^{\prime}\to S_{\Sigma}}\int_{S^{\prime}} \frac{-\tilde \varepsilon_{2}}{2}\, \Sigma^{-2}\, \vartheta(K^{a})\, \mc{S}_{c}\, \varepsilon^{c}{}_{b} X^{b}\notag\\
    & -\lim_{S^{\prime} \to S_{\Sigma}}\,\frac{1}{2} \int_{S^{\prime}} \tilde \varepsilon_{2}\, K^{a} \nabla_{a} (\Omega^{-1}\,\Sigma^{-2}\, \ast C_{bcde}\,n^{d}\,n^{b}\,l^{c}\,X^{e}) = \int_{ S_{\Sigma}} \tilde{\varepsilon}_{2}\, \big[ \Sigma^{-2} X^{b} \varepsilon_{b}{}^{a} (\mc{J}_{a} + \frac{1}{4} \ms{D}_{a} \mc{P}) + \cdots \big]\,,
\end{align}
where we used $\vartheta(K^{a}) \hateq \vartheta(l^{a})$ to simplify the final expression. As shown in \cref{eq:isfinite}, the right hand side of \cref{maineq-3-2} is finite in the limit $\Sigma^{-1} \to 0$. As a result, the limit of \cref{eq:beta-component} to $\nulls^{+}$ along $\scrp$ gives 
  \be \label{eq:finaleq}
 -\frac{1}{8\pi} \int_{\nulls^+}\,\tilde \varepsilon_{2} \big[\Sigma^{-2} X^{b}\,\varepsilon_{b}{}^{a} (\mc{J}_{a} + \frac{1}{4} \ms{D}_{a} \mc{P})\big]\,,
  \ee
  which, from \cref{eq:Q-GR-on-N+}, corresponds to the limit of the charge associated with any Lorentz symmetry given by $\tilde{\varepsilon}_{b}{}^{a} X^{b} $ (using the fact that $\tilde{\varepsilon}_{b}{}^{a} = \varepsilon_{b}{}^{a}$). In the same way, the limit of \cref{eq:beta-component} to $\nulls^{-}$ along $\scri^{-}$ also yields
  \be
   -\frac{1}{8\pi} \int_{\nulls^-}\,\tilde \varepsilon_{2} \big[\Sigma^{-2} X^{b}\,\varepsilon_{b}{}^{a} (\mc{J}_{a} + \frac{1}{4} \ms{D}_{a} \mc{P})\big]\,.
    \ee
Therefore, assuming continuity of \cref{eq:beta-component} at $\nulls^{\pm}$, Lorentz charges, in the limit to $\nulls^\pm$ along $\scri^{\pm}$, match the Lorentz charges in the limit to $\nulls^\pm$ along $\cyl$. Since, as discussed earlier, the Lorentz charges (in Ashtekar-Hansen gauge) on $\hyp$ are conserved and therefore their values on the $\nulls^{\pm}$ are the same, it follows that the Lorentz charges on $\scri^{+}$ and $\scri^-$ match in the limit to spatial infinity.  

Using the proof of matching of supertranslation symmetries and the associated supermomentum charges in \cite{KP-GR-match}, we then conclude that \emph{all} BMS symmetries on past and future null infinity match antipodally in the limit to spatial infinity and their charges become equal in this limit. This immediately implies the following infinitely many conservation laws (one for each pair of ``matched'' generators $\xi^{a+}$ and $\xi^{a-}$)
 \begin{equation} \label{eq:cons-law}
     \mathcal{Q}\left[\xi^{a +} ; S_{\Sigma}^{+}\right]-\mathcal{Q}\left[\xi^{a -} ; S_{\Sigma}^{-}\right]=\mathcal{F}^{+}\left[\xi^{a +}; \Delta\mathscr{I}^{+}\right]+\mathcal{F}^{-}\left[\xi^{a -}; \Delta\mathscr{I}^{-}\right]\,,
     \end{equation}
where $\xi^{a+}$ denotes a BMS symmetry on $\scri^+$, $\xi^{a-}$ denotes the BMS symmetry on $\scri^-$ that this matches onto in the limit to spatial infinity, $\mathcal{F}^{+}$ ($\mathcal{F}^{-}$) denotes the incoming flux\footnote{In the orientation conventions picked in this paper, as in \cite{KP-EM-match,KP-GR-match}, the fluxes at both $\scri^{\pm}$ are incoming.} of charge associated with a BMS symmetry on $\scri^+$ ($\scri^{-}$), $\Delta \scri^{+}$ ($\Delta \scri^{-}$) denotes a portion of $\scri^+$ ($\scri^-$) between spatial infinity and a cross-section, $S_{\Sigma}^{+}$ ($S_{\Sigma}^{-}$), of future (past) null infinity. As discussed in \cite{KP-GR-match}, if suitable fall-off conditions are satisfied on the future and past boundaries of future and past null infinity (i.e in the limit to timelike infinities) such that the BMS charges go to zero in those limits then we obtain the global conservation law
\begin{equation} \label{eq:cons-law-2}
\mathcal{F}^{+}\left[\xi^{a +}; \Delta\mathscr{I}^{+}\right]+\mathcal{F}^{-}\left[\xi^{a -}; \Delta\mathscr{I}^{-}\right]=0\,,
 \end{equation}
that is, the total incoming flux equals the total outgoing flux for all BMS symmetries and therefore that the flux is conserved in any classical gravitational scattering process from $\scri^{-}$ to $\scri^{+}$.

\section{Discussion}
\label{sec:disc}

We showed the antipodal matching of Lorentz symmetries and the equality of the associated charges on past and future null infinity in the limit to spatial infinity in a class of spacetimes that are asymptotically-flat at null and spatial infinity in the sense of Asthekar and Hansen. Combined with the result of \cite{KP-GR-match} where the matching of supertranslation symmetries and supermomentum charges was similarly shown, this proves the matching of \emph{all} BMS symmetries and charges in these spacetimes. While we did not require that our spacetimes be stationary, we did make the following assumptions about our class of spacetimes:
\begin{enumerate*}
\item  we assumed that at null infinity the peeling theorem holds;\footnote{We only need enough regularity at null infinity so that the Weyl tensor components \(\mc R_{ab}, \mc S_a, \mc P, \mc P^*, \mc J_a\) are defined.}
\item that $\dd{B}_{ab} =0$, which, as discussed earlier, is known to be true (at least) in asymptotically-flat spacetimes that are \emph{either} stationary \emph{or} axisymmetric; \item that $\dd{\beta}_{ab}$ is odd under the reflection map on $\hyp$;
\item that the spacetimes are null-regular at $i^{0}$ in the sense of  \cref{eq:P-falloff,eq:N-R-falloff} as assumed in \cite{KP-GR-match};
\item that the trace-free projection of (rescaled) $S_{ab}$ is continuous at $\nulls^{\pm}$ (in the precise sense discussed in \cref{sec:fixing-st}); and
\item that the (rescaled) Weyl tensor component (appearing in \cref{eq:beta-component}) is continuous at $\nulls^{\pm}$.
\end{enumerate*}
These conditions are sufficient (but possibly not necessary) to prove the matching of the BMS charges at spatial infinity.

To confirm the viability of these conditions, one can check them for explicit solutions like the Kerr-Newman family of spacetimes. In this case, following \cite{Herb-Kerr}, one can construct a conformal-completion where the unphysical metric is \(C^{>0}\) at spatial infinity along both null and spatial directions. The smoothness conditions at null infinity are automatically satisfied, and one can explicitly check that all our regularity conditions at spatial infinity are also satisfied. Further, these conditions also hold if one \emph{assumes} that the unphysical spacetime is \(C^{>1}\) and the unphysical metric is \(C^{>0}\) at spatial infinity in both null and spatial directions (as has implicitly been assumed in \cite{Capone:2022gme}).

The validity of these conditions for general solutions of the Einstein equations remains an open problem. To definitively settle the validity of these conditions, one would have to show that there exists a ``large enough'' class of initial data --- with some suitable topology --- on a Cauchy surface or on \(\scri^-\) which evolve under the Einstein equations to spacetimes which satisfy the conditions we have imposed. The Ashtekar-Hansen structure (\cref{def:AH}) is motivated by the construction of asymptotically-flat initial data at spatial infinity on a Cauchy surface by Geroch \cite{Geroch-asymp}. However, this formulation does not easily allow us to evolve the data all the way to null infinity. The evolution of initial data to null infinity would be most naturally addressed in the formalism of Friedrich \cite{Friedrich} (see \cite{Tro,Mohamed:2021rfg} for the analysis in linearized gravity around Minkowski spacetime). Of particular interest would be polyhomogenous spacetimes at null infinity \cite{polyhom}, where the peeling theorem is not satisfied and \(\mc J_a\) does not exist at null infinity. Similarly, the definition of the Lorentz charges at spatial infinity must be modified if \(\dd B_{ab}\neq 0\) (see \cite{CD,CDV-Lorentz,Prabhu:2019daz}) or if the Ashtekar-Hansen falloff conditions at spatial infinity along spatial directions do not hold \cite{Herb-dd}. It would be of interest to investigate the matching of Lorentz charges in these cases. 

It would also be interesting to investigate matching conditions between symmetries and charges defined on black hole horizons \cite{CFP} and those associated with null infinity. This would require analyzing symmetries and charges in the limit to timelike infinities using the framework developed in \cite{HL-timelike}. We leave a detailed study of this to future work.

\section*{Acknowledgements}

We thank \'Eanna \'E. Flanagan for some key insights and constant encouragement over the course of this work. This work was supported in part by the NSF grants PHY-1707800 and PHY-2110463 to Cornell University. Additionally, I.S. is supported in part by the John and David Boochever prize fellowship in fundamental theoretical physics. K.P. is supported in part by  NSF grants PHY-1801805 and PHY-2107939. Some calculations in this paper used the computer algebra system \textsc{Mathematica} \cite{Mathematica}, in combination with the \textsc{xAct/xTensor} suite~\cite{JMM:xAct,MARTINGARCIA2008597}, and the Riemannian Geometry and Tensor Calculus package \cite{RGTC}.

\appendix

\section{Spin-weighted spherical harmonics}
\label{sec:swsh}

In this appendix, we include some results on stereographic coordinates, spin transformations and spin-weighted spherical harmonics that are used in the body of the paper. 
 
\paragraph*{Stereographic coordinates:}Stereographic coordinates \(\theta^A = ( z, \bar z)\) are related to the usual spherical polar coordinates \((\theta,\phi)\) by
\be \label{eq:stereocoords-defn}
    z = e^{i\phi} \cot \nfrac{\theta}{2} \eqsp \bar{z} = e^{-i\phi} \cot \nfrac{\theta}{2}\,. 
\ee
The unit round sphere metric and the area-element in these coordinates are given by
\begin{subequations}
\be \label{eq:stereographic-metric}
    s_{AB} \equiv 2 P^{-2} d z d\bar z \,,
    \ee
    \be \label{eq:stereographic-areael}
    \varepsilon_{AB} \equiv i P^{-2} dz \wedge d\bar z = -\sin \theta d\theta \wedge d\phi \,,
\ee
\end{subequations}
where
\be\label{eq:P-defn}
    P \defn \frac{1+ z \bar z}{\sqrt{2}}\,.
\ee
The antipodal map is implemented by the transformation
\be \label{eq:antipodalmap}
    \theta^A \mapsto - \theta^A : z \mapsto - 1/\bar z \implies P^{-1} dz \mapsto e^{2 i \phi} P^{-1} d\bar z \,,
\ee
where \(e^{2 i\phi} = z/\bar z\). 

\paragraph*{Spin transformations:}A function \(\eta\) is said to have spin-weight \(s\) if
\be \label{eq:spin-wt-defn}
    P^{-1} dz \mapsto e^{i\lambda} P^{-1} dz \implies \eta \mapsto e^{i s \lambda} \eta\,,
\ee
where \(\lambda\) is any smooth function on \(\bb S^2\). We denote this as $\eta \wt s$. On spin-weighted functions, we defined the operators \(\eth\) and \(\bar\eth\) by 
\be \label{eq:ethethbaraction}
    \eth \eta \defn P^{1-s} \frac{\partial}{\partial\bar z} (P^s \eta)  \eqsp \bar\eth \eta \defn P^{1+s} \frac{\partial}{\partial z} (P^{-s} \eta)\,.
\ee
Note that if \(\eta\) has spin-weight \(s\) then \(\eth\eta\) has spin-weight \(s+1\) and \(\bar\eth \eta\) has spin weight \(s-1\). Further, we have the relations
\be
    (\eth\bar\eth - \bar\eth \eth) \eta = -s \eta  \eqsp \ms D^2\eta = ( \eth \bar\eth + \bar\eth \eth ) \eta \,,
\ee
where \(\ms D^2\) is the Laplacian corresponding to the unit round metric on \(\bb S^2\).

\paragraph*{Spin-weighted spherical harmonics:}The \emph{spin-weighted spherical harmonics} \(Y_{\ell,m}^s(\theta^A)\) are eigenfunctions of the Laplacian with spin-weight \(s\). They satisfy \cite{PR1,Stewart}
\be\label{eq:swsh-laplacian}
    \ms D^2 Y_{\ell,m}^s = - \lb[ \ell(\ell+1) - s^2 \rb] Y_{\ell,m}^s\,.
\ee
\be\label{eq:swsh-eth}
    \eth Y_{\ell,m}^s = - \sqrt{\frac{(\ell-s)(\ell+s+1)}{2}} Y_{\ell,m}^{s+1} \eqsp \bar\eth Y_{\ell,m}^s = \sqrt{\frac{(\ell+s)(\ell-s+1)}{2}} Y_{\ell,m}^{s-1}\,.
\ee
Note that the harmonic \(Y_{\ell,m}^s\) is non-vanishing only for \(\ell \geq |s|\) and \(\ell \geq |m|\). An explicit expression for \(Y_{\ell,m}^s\) as functions of the coordinates \((z,\bar z)\) is given in Eq.~3.9.20 of \cite{Stewart}.

 Further, under complex conjugation and antipodal reflection we have
\begin{subequations}\begin{align}
     \bar{Y_{\ell,m}^s}(\theta^A) & = (-)^{m+s} Y_{\ell,-m}^{-s}(\theta^A) \label{eq:swsh-cc}\,,\\
     Y_{\ell, m}^s(-\theta^A) & = (-)^{\ell + s} e^{2 i s \phi}Y_{\ell,m}^{-s}(\theta^A) \label{eq:swsh-parity}\,.
\end{align}\end{subequations}

\section{Solutions for reflection-odd \(\dd \beta_{ab}\)}
\label{sec:beta-soln}
As discussed in the body of the paper, we restrict ourselves to solutions for $\dd{\beta}_{ab}$ that are odd under the reflection map defined in \cref{eq:reflection-hyp}. In this appendix, we solve the equations of motion that govern these solutions, that is, \cref{eq:div-beta-2,eq:waveeqn-beta}, and study the limits of the solutions to $\nulls^{\pm}$. As in \cref{sec:LorentzsymmonH}, we make use of $(\alpha, z, \bar{z})$ coordinates in which the metric is given by \cref{eq:hyp-metric} and the limits to $\nulls^{\pm}$ correspond to $\alpha \to \pm 1$. Using the fact that $\dd{\beta}_{ab}$ is traceless, its components can be written as
\be \label{eq:beta-comp-def}
    \dd\beta_{\alpha \alpha} = H \eqsp \dd\beta_{\alpha  z} = P^{-1} J \eqsp \dd\beta_{ z  z} = P^{-2} G \eqsp \dd\beta_{ z \bar z} = \tfrac{1}{2} P^{-2} (1-\alpha^2) H \,,
\ee
where
\be
    H\wt s = 0 \eqsp J \wt s = -1 \eqsp G \wt s = -2 \,,
\ee
Note that the condition that $\dd{\beta}_{ab}$ be reflection-odd gives us the following conditions
\be \label{eq:parity-1}
    H(-\alpha,-\theta^A) = - H(\alpha,\theta^A) \eqsp J(-\alpha,-\theta^A) = e^{-2 i \phi}\bar J(\alpha,\theta^A) \eqsp G(-\alpha,-\theta^A) = - e^{-4 i \phi}\bar G(\alpha,\theta^A) \,.
\ee
In terms of these components, \(\dd D^b \dd\beta_{ab} = 0\) corresponds to the following equations
\begin{subequations}\label{eq:div-beta-coord}\begin{align}
    0 & = (1-\alpha^2) \partial_\alpha H - \alpha H - \eth J - \bar{\eth}\, \bar J \label{eq:div-H-coord} \,, \\
    0 & = (1-\alpha^2) \partial_\alpha J - \tfrac{1}{2} (1-\alpha^2) \bar\eth H - \eth G \label{eq:div-J-coord} \,,
\end{align}\end{subequations}
while \((\dd D^2 - 2) \dd\beta_{ab} = 0\) corresponds to
\begin{subequations}\label{eq:wave-beta-coord}\begin{align}
    0 & = (1-\alpha^2) \partial_\alpha^2 H - 4 \alpha \partial_\alpha H - (\ms D^2 + 2 )H \,,\\
    0 & = (1-\alpha^2) \partial_\alpha^2 J - 4 \alpha \partial_\alpha J - (\ms D^2 +1 ) J +  2 \alpha \bar\eth H \,, \\
    0 & = (1-\alpha^2) \partial_\alpha^2 G - 4 \alpha \partial_\alpha G - \ms D^2 G + 4 \alpha \bar\eth J \,,
\end{align}\end{subequations}
Note also that to simplify the first and second equations above, we used \cref{eq:div-H-coord} and \cref{eq:div-J-coord} respectively.
Expanded in terms of spin-weighted spherical harmonics, we have
\be
    H = \sum_{\ell ,m} H_{\ell,m}(\alpha) Y_{\ell,m}^{s=0} (\theta^A) \eqsp
    J = \sum_{\ell \geq 1,m} J_{\ell,m}(\alpha) Y_{\ell,m}^{s=-1} (\theta^A) \eqsp
    G = \sum_{\ell \geq 2,m} G_{\ell,m}(\alpha) Y_{\ell,m}^{s=-2}(\theta^A) \,.
\ee
The conditions for the solutions to be reflection-odd in \cref{eq:parity-1} then imply the following relations
\be \label{eq:parity-2}
H_{\ell,m} (-\alpha) = (-1)^{\ell+1} H_{\ell,m}(\alpha)\,, \, J_{\ell,m}(-\alpha) =(-1)^{\ell+m} \bar{J_{\ell,-m}}(\alpha)\,, \, G_{\ell,m}(-\alpha) = (-1)^{1+\ell+m} \bar{G_{\ell,-m}} (\alpha)\,.
\ee
Note also that since $H$ is real, it follows from \cref{eq:swsh-cc} that
\be \label{eq:Hlm-reality}
H_{\ell,m} = (-1)^m \bar{ H_{\ell,-m}}\,.
\ee
Then, \cref{eq:div-beta-coord} becomes
\begin{subequations}\label{eq:div-beta-harmonics}\begin{align}
    0 & = (1-\alpha^2) \frac{d}{d\alpha} H_{\ell,m} - \alpha H_{\ell,m} + \sqrt{\tfrac{\ell (\ell+1)}{2}} ( J_{\ell,m} + (-1)^{m} \bar{J_{\ell,-m}}) \label{eq:div-H-harmonics} \,.\\
    0 & = (1-\alpha^2) \frac{d}{d\alpha} J_{\ell,m} - \tfrac{1}{2} (1-\alpha^2) \sqrt{\tfrac{\ell (\ell+1)}{2}} H_{\ell,m} +  \sqrt{\tfrac{(\ell - 1) (\ell+2)}{2}} G_{\ell,m}\,. \label{eq:div-J-harmonics}
\end{align}\end{subequations}
Moreover, using \cref{eq:swsh-laplacian} and \cref{eq:Hlm-reality}, \cref{eq:wave-beta-coord} becomes
\begin{subequations}\label{eq:wave-beta-harmonics}\begin{align}
    0 & = (1-\alpha^2) \frac{d^2}{d\alpha^2} H_{\ell,m} - 4 \alpha \frac{d}{d\alpha} H_{\ell,m} + \lb[ \ell(\ell+1) - 2 \rb] H_{\ell, m} \label{eq:H-eq} \,, \\
    0 & = (1-\alpha^2) \frac{d^2}{d\alpha^2} J_{\ell,m} - 4 \alpha \frac{d}{d\alpha} J_{\ell,m} + \lb[ \ell(\ell+1) -2 \rb] J_{\ell,m} + 2 \alpha \sqrt{\tfrac{\ell (\ell+1)}{2}} H_{\ell,m} \label{eq:J-eq}\,,  \\
    0 & = (1-\alpha^2) \frac{d^2}{d\alpha^2} \bar{J_{\ell,m}} - 4 \alpha \frac{d}{d\alpha} \bar{J_{\ell,m}} + \lb[ \ell(\ell+1) -2 \rb] \bar{J_{\ell,m}} + 2(-1)^{m} \alpha \sqrt{\tfrac{\ell (\ell+1)}{2}} H_{\ell,-m} \label{eq:Jbar-eq}\,, \\
    0 & = (1-\alpha^2) \frac{d^2}{d\alpha^2} G_{\ell,m} - 4\alpha \frac{d}{d\alpha} G_{\ell,m} + \lb[\ell(\ell+1) - 4\rb] G_{\ell,m} + 4 \alpha \sqrt{\tfrac{(\ell - 1) (\ell+2)}{2}} J_{\ell,m}\,, \label{eq:G-eq}
\end{align}\end{subequations}
where in each of these equations, we have suppressed the dependence of $H_{\ell,m}$, $J_{\ell,m}$ and $G_{\ell,m}$ on $\alpha$. Note also that \cref{eq:Jbar-eq} was obtained by taking the complex conjugate of \cref{eq:J-eq} and using \cref{eq:Hlm-reality}.

We now proceed to solving these equations, starting first by deriving explicit solutions for the $\ell=0$ and $\ell=1$ cases. Note that $G=0$ in both these cases.  For \(\ell=0\), \(J =0\) and we have the solution
\be
 H_{\ell=0,m=0}(\alpha) = (1-\alpha^2)^{-1} (H_0 + H_1 \alpha)\,,
 \ee
  which satisfies \cref{eq:div-H-harmonics} only for \(H_0 = H_1 = 0\). Therefore all \(\ell=0\) solutions are zero.

For \(\ell = 1\), we have the solutions 
\be\begin{aligned}
    H_{\ell=1,m}(\alpha) & = H_m^{(0)} + H_m^{(1)} (\alpha (1-\alpha^2)^{-1} + \tanh^{-1} \alpha)\,, \\
    J_{\ell=1,m}(\alpha) & = J_m^{(0)} + J_m^{(1)} \alpha \,.\\
\end{aligned}\ee
Putting these in \cref{eq:div-beta-harmonics} gives us the conditions
\be\begin{aligned}
     H^{(1)}_m = 0 \eqsp J^{(0)}_m + (-)^m \bar {J^{(0)}_{-m}} = 0 \eqsp H^{(0)}_m = J^{(1)}_m + (-)^m \bar {J^{(1)}_{-m}} \eqsp H^{(0)}_m = 2 J^{(1)}_m \,,
\end{aligned}\ee
and the last two equations together imply
\be
    J^{(1)}_m - (-)^m \bar {J^{(1)}_{-m}} = 0 \,.
\ee

With these constraints, the free functions correspond to 6 degrees of freedom 
\be \label{eq:Jdof}
    J^{(0)}_{m=1} \in \bb C \eqsp i J^{(0)}_{m=0} \in \bb R \eqsp J^{(1)}_{m=1} \in \bb C \eqsp J^{(1)}_{m=0} \in \bb R \,.
\ee
We will see below that these encode the 6 charges associated with Lorentz boosts and rotations. Note that we have shown that the $\ell=0,1$ solutions are finite in the limits $\alpha \to \pm1$. We now proceed to showing that this finiteness property holds for all $\ell$. Note that the solutions of \cref{eq:H-eq} for $\ell \geq 1$ are spanned by
\be\begin{aligned}
    (1-\alpha^2)^{-\half} P_\ell^{-1}(\alpha) \eqsp (1-\alpha^2)^{-\half} Q_\ell^{-1}(\alpha)\, \\
 \end{aligned}\ee
where $P_{\ell}^{-1} (\alpha)$ and $Q_{\ell}^{-1} (\alpha)$ are the Legendre functions, which satisfy
\be \label{eq:PQparity}
P_{\ell}^{-1} (-\alpha) = (-1)^{\ell-1} P_{\ell}^{-1} (\alpha)  \, \eqsp Q_{\ell}^{-1} (-\alpha) = (-1)^{\ell} Q_{\ell}^{-1} (\alpha) \,.
\ee
We see that reflection-odd condition on $H_{\ell,m}$ in \cref{eq:parity-2} picks out the solutions spanned by $(1-\alpha^{2})^{-\half} P^{-1}_{\ell}(\alpha)$ for $H_{\ell,m}$.
Hence, we have that
\be \label{eq:Hhigherlsoln}
H_{\ell>1,m}(\alpha) = c_{m} (1-\alpha^{2})^{-\half}P^{-1}_{\ell}(\alpha)\,,
\ee
for some constants $c_m$.
 It follows from the following recursion relation 
\be \label{eq:recursion}
(1-\alpha^2)^{-\half} P_\ell^{-1}(\alpha) = \frac{-1}{2^{\ell} \ell! (1-\alpha^{2})} \big(\frac{d}{d\alpha} \big)^{\ell-1} (\alpha^{2}-1)^{\ell} \,\,\,\, \text{for $\ell>1$} \,,
\ee 
that these solutions are finite as $\alpha \to \pm 1$ for all $\ell>1$. 
Note also that \cref{eq:J-eq,eq:Jbar-eq} can be combined to give
\begin{align}
0 = (1-\alpha^2) \frac{d^2}{d\alpha^2} (J_{\ell,m} -(-1)^{m} \bar{J_{\ell,-m}}) &- 4 \alpha \frac{d}{d\alpha} (J_{\ell,m} -(-1)^{m} \bar{J_{\ell,-m}}) \notag \\
& + \lb[ \ell(\ell+1) - 2 \rb] (J_{\ell,m} -(-1)^{m} \, \bar{J_{\ell,-m}})
\end{align}
This shows that $J_{\ell,m}(\alpha) -(-1)^{m} \bar{J_{\ell,-m}}(\alpha)$ satisfies the same equation as $H_{\ell,m}$ and therefore the solutions to this equation are also spanned by linear combinations of 
\be \label{eq:J-higher-l-soln}
    (1-\alpha^2)^{-\half} P_\ell^{-1}(\alpha) \eqsp (1-\alpha^2)^{-\half} Q_\ell^{-1}(\alpha)\,.
    \ee
Using the reflection-odd condition on $J_{\ell,m}$ given in \cref{eq:parity-2}, we see that $J_{\ell,m}(\alpha) -(-1)^{m} \bar{J_{\ell,-m}}(\alpha) = J_{\ell,m}(\alpha) - (-1)^{\ell} J_{\ell,m}(-\alpha)$ which is odd (even) under $\alpha \to -\alpha$ for even (odd) $\ell$. From \cref{eq:PQparity}, we see that this property only holds for the solutions spanned by $ (1-\alpha^2)^{-\half} P_\ell^{-1}(\alpha)$ which leads us to discard the solutions spanned by $(1-\alpha^2)^{-\half} Q_\ell^{-1}(\alpha)$. Hence, we have
\be \label{eq:J-higher-l-soln2}
J_{\ell>1,m}(\alpha) -(-1)^{m} \bar{J_{\ell>1,-m}}(\alpha)  = d_{m} (1-\alpha^{2})^{-\half} P^{-1}_{\ell}(\alpha)\,,
\ee
for some constants $d_m$. Using \cref{eq:recursion} again, we conclude that the solutions for  $J_{\ell,m} -(-1)^{m} \bar{J_{\ell,-m}}$ are finite as $\alpha \to \pm 1$ for all $\ell$. Next, note that we can rewrite \cref{eq:div-H-harmonics} as
\be \label{eq:JmmJmm}
  J_{\ell,m} + (-1)^{m} \bar{J_{\ell,-m}} =\sqrt{\frac{2}{\ell(\ell+1)}} \bigg[ (\alpha^2-1) \frac{d}{d\alpha} H_{\ell,m} + \alpha H_{\ell,m}\bigg] \,\,\,\, \text{(for $\ell \geq 1$)}\,.
 \ee
Using \cref{eq:recursion}, we see that $(\alpha^{2} -1) \frac{d}{d\alpha} H_{\ell,m}$ is finite as $\alpha \to \pm 1$ which, using \cref{eq:JmmJmm} and the finiteness of $H_{\ell,m}$ in these limits, implies that $J_{\ell,m} + (-1)^{m} \bar{J_{\ell,-m}}$ is also finite as $\alpha \to \pm 1$ for all $\ell \geq 1$. Using the finiteness of $J_{\ell,m} - (-1)^{m} \bar{J_{\ell,-m}}$ for all $\ell$ in the limit $\alpha \to \pm 1$, we then conclude that $J_{\ell,m}(\alpha)$ is finite as $\alpha \to \pm 1$ for all $\ell \geq 1$.

Finally, we note from \cref{eq:div-J-harmonics} that
\be
G_{\ell,m}  = \sqrt{\frac{2}{(\ell-1)(\ell+2)}} \bigg[(\alpha^2-1) \frac{d}{d\alpha} J_{\ell,m} -\tfrac{1}{2} (\alpha^2-1) \sqrt{\tfrac{\ell (\ell+1)}{2}} H_{\ell,m} \bigg] \,\,\,\, \text{(for $\ell \geq 2$)} \,.
\ee
Using \cref{eq:JmmJmm,eq:J-higher-l-soln2} to obtain the functional behavior of $J_{\ell,m}(\alpha)$ and using \cref{eq:recursion}, one can also show that $(\alpha^2-1) \frac{d}{d\alpha} J_{\ell,m}$ is finite as $\alpha \to \pm 1$. This, using the finiteness of $H_{\ell,m}$ in these limits, shows that $G_{\ell,m}$ is finite as $\alpha \to \pm 1$ for all $\ell \geq 2$. This completes our proof of the fact that the reflection-odd solutions for $\dd{\beta}_{ab}$ are finite in the $\alpha \to \pm 1$ limits for all $\ell$.

One can show that in the coordinates used in this section, $\lim_{\to \nulls^{\pm}} \dd{\Sigma}^{-1} \dd{U}^{a}|_{\nulls^{\pm}} \equiv \pm \partial_{\alpha}$. We therefore see from \cref{eq:AHchargeonnulls} that the (integrand of) Lorentz charge at $\nulls^{\pm}$ is proportional to $\dd{\beta}_{\alpha A} \dd{X}^{A}$ where $A=(z,\bar{z})$. Since $X^{A}$ (as shown in \cref{sec:LorentzsymmonH}) is an $\ell=1$ vector field, it follows that the charge receives contribution only from $J_{\ell=1,m}$. The six degrees of freedom of $J_{\ell=1,m}$ (\cref{eq:Jdof}) therefore encode the six Lorentz charges associated with boosts and rotations.

\section{Affinity of $l^{a}$}

For the calculation in \cref{sec:matching}, we need the expression for the affinity of $l^{a}$. Recall that we are working in a conformal frame where $\Phi=2$ and so from \cref{eq:n-Phi} we have
\be \label{eq:derninphi2}
\nabla_{a} n_{b} \hateq 2 g_{ab}\,.
\ee
Using \cref{eq:l-defn}, we have 
\be \label{eq:affinity-l}
l^{a} \nabla_{a} l_{b}=\frac{1}{4} \Sigma L^{a} (\Sigma \nabla_{a}  L_{b} + L_{b} \nabla_{a} \Sigma)\,.
\ee
   
Then, using \cref{eq:Lamodified,eq:N-defn} and \cref{eq:derninphi2}, we get 
\begin{align}
\nabla_{a} L_{b} & \hateq - \nabla_{a} \nabla_{b} \Sigma^{-1} + \tfrac{1}{2} \Sigma n_{b} \nabla_{a} \nabla_{c} \Sigma^{-1} \nabla^{c} \Sigma^{-1} + \tfrac{1}{4} \nabla_{c} \Sigma^{-1} \nabla^{c} \Sigma^{-1}  (n_{b} \nabla_{a} \Sigma + 2 \Sigma   g_{ab} ) \notag\\
    &\quad~ - \tfrac{1}{2} \nabla_{a} (\Omega \Sigma \bar{L}^{c} \nabla_{b} \nabla_{c} \Sigma^{-1})\,,
\end{align}
which gives
\begin{align} \label{eq:geo-l}
    l^{a} \nabla_{a} l_{b} & \hateq  \tfrac{1}{4} \Sigma L_{b} L^{a} \nabla_{a} \Sigma + \tfrac{1}{4} \Sigma^{2} L^{a} \big[- \nabla_{a} \nabla_{b} \Sigma^{-1} + \tfrac{1}{2} \Sigma n_{b} \nabla_{a} \nabla_{c} \Sigma^{-1} \nabla^{c} \Sigma^{-1} \notag\\
    &\qquad + \tfrac{1}{4} \nabla_{c} \Sigma^{-1} \nabla^{c} \Sigma^{-1}  (n_{b} \nabla_{a} \Sigma + 2 \Sigma  g_{ab} )\big] - \tfrac{1}{8} \Sigma^{2} L^{a} \nabla_{a} ( \Omega \Sigma \bar{L}^{c} \nabla_{b} \nabla_{c}\Sigma^{-1})\,.
\end{align}
Let us use this to compute, $q^{b}{}_{c}\, l^{a} \nabla_{a} l_{b}$ on $\scri$. Using $q^{a}{}_{b}\, n_{a} \hateq q^{a}{}_{b}\, L_{a} \hateq 0$ and $L^{a} \nabla_{a} \Omega \hateq -2 \Sigma^{-1}$ (where the last equation follows from \cref{eq:NL-LL,eq:N-defn}), we have      
  \be \label{eq:q-proj-affinity}
   q^{b}_{c}\, l^{a} \nabla_{a} l_{b} \hateq - \frac{1}{4} \Sigma^{2} q^{b}{}_{c}  L^{a} \nabla_{a} \nabla_{b} \Sigma^{-1} + \frac{1}{4} \Sigma^{2} q^{b}{}_{c} \bar{L}^{a} \nabla_{a} \nabla_{b}\Sigma^{-1} \hateq 0\,,
   \ee           
   where the last equality follows because $L^{a} \hateq \bar{L}^{a}$ (see \cref{eq:barLa}).
Next, we evaluate $n^{b} l^{a} \nabla_{a} l^{b}$ on $\scri$. Using \cref{eq:affinity-l,eq:N-defn,eq:NL-LL}, we have
\begin{align}
n^{b} l^{a} \nabla_{a} l_{b} & =\frac{1}{4} \Sigma^{2} L^{a} n^{b} \nabla_{a} L_{b} + \frac{1}{4} \Sigma L^{a}  n^{b} L_{b} \nabla_{a} \Sigma \notag\\
          &\hateq \frac{1}{2} \Sigma L^{a} N^{b} \nabla_{a} L_{b} - \frac{1}{2} L^{a} \nabla_{a} \Sigma \,.
\end{align}
Note that using \cref{eq:Lamodified} 
\begin{align}
L^{a} \nabla_{a} \Sigma & =-\nabla^{a} \Sigma^{-1} \nabla_{a} \Sigma + \frac{1}{2} N^{a} \nabla_{a} \Sigma \nabla_{c} \Sigma^{-1} \nabla^{c} \Sigma^{-1}- \frac{1}{2} \Omega \Sigma \bar{L}^{b} \nabla^{a} \nabla_{b} \Sigma^{-1} \nabla_{a} \Sigma\notag\\
          & \hateq \frac{1}{2} \Sigma^{2} \nabla_{a} \Sigma^{-1} \nabla^{a} \Sigma^{-1}\,,
\end{align}
where the second equality uses $N^{a} \nabla_{a} \Sigma^{-1} \hateq 1$ (which follows from \cref{cond:Sigma-choice,eq:N-defn}) and the fact that $\Omega \hateq 0$. Moreover,
\begin{align}
L^{a} N^{b} \nabla_{a} L_{b} &= -L^{a} L_{b} \nabla_{a} N^{b} + L^{a} \nabla_{a} (N^{b} L_{b}) \hateq  - L^{a} L_{b} (-\Sigma N^{b} \nabla_{a} \Sigma^{-1} +  \Sigma\, \delta^{b}{}_{a}) + L^{a} \nabla_{a} (N^{b} L_{b})\notag\\
&\hateq - L^{a}\Sigma \nabla_{a}\Sigma^{-1} + L^{a} \nabla_{a} (-N^{b} \nabla_{b} \Sigma^{-1} + \frac{1}{2}N^{b} N_{b} \nabla_{c} \Sigma^{-1}\nabla^{c} \Sigma^{-1} -\frac{1}{2} \Omega \Sigma \bar{L}^{b} N^{c} \nabla_{c} \nabla_{b} \Sigma^{-1})\,,
\end{align}
 where the first line uses \cref{eq:N-defn} and \cref{eq:derninphi2} while the second line uses \cref{eq:NL-LL,eq:Lamodified}. Further, using $N^{a} N_{a} = \Sigma^{2} \Omega + O(\Omega^{2})$ (which follows from \cref{eq:N-defn} and the fact that $\lim_{\to \scri} \Omega^{-1} n^{a} n_{a}= 2\Phi= 4$), and $L^{a} \nabla_{a} \Omega \hateq -2 \Sigma^{-1}$, this becomes
\begin{align}
        L^{a} N^{b} \nabla_{a} L_{b}    &\hateq - L^{a}\Sigma \nabla_{a}\Sigma^{-1} + L^{a} \nabla_{a} (-N^{b} \nabla_{b} \Sigma^{-1} + \frac{1}{2} \Sigma^{2} \Omega\nabla_{b} \Sigma^{-1}\nabla^{b} \Sigma^{-1}) +\bar{L}^{a} N^{b} \nabla_{b} \nabla_{a} \Sigma^{-1}\notag\\
        &\hateq- L^{a} \Sigma \nabla_{a} \Sigma^{-1} - L^{a} \nabla_{a} (N^{b} \nabla_{b} \Sigma^{-1}) -\Sigma \nabla_{a} \Sigma^{-1}\nabla^{a} \Sigma^{-1}+ \bar{L}^{a} N^{b} \nabla_{b} \nabla_{a} \Sigma^{-1}\,,
        \end{align}                  
Putting all of this together, we have
\begin{align}
    n^{b} l^{a} \nabla_{a} l_{b}
    &\hateq \frac{1}{2} \Sigma L^{a} N^{b} \nabla_{a} L_{b} - \frac{1}{2} L^{a} \nabla_{a} \Sigma \notag\\
    &\hateq -\frac{1}{4} \Sigma^{2} \nabla_{a} \Sigma^{-1} \nabla^{a} \Sigma^{-1} + \frac{1}{2} (-L^{a} \Sigma^{2} \nabla_{a} \Sigma^{-1}-\Sigma L^{a} \nabla_{a} (N^{b} \nabla_{b} \Sigma^{-1}) - \Sigma^{2} \nabla_{a} \Sigma^{-1} \nabla^{a}\Sigma^{-1})\notag\\
    &\qquad +\frac{1}{2} \Sigma \bar{L}^{a} N^{b} \nabla_{b} \nabla_{a} \Sigma^{-1}\,.
    \end{align}
Note that using \cref{eq:Lamodified} and $N^{a} \nabla_{a} \Sigma^{-1}\hateq 1$, we have 
\be \label{eq:l-der-sigmainv}
L^{a} \nabla_{a} \Sigma^{-1} \hateq - \nabla^{a} \Sigma^{-1} \nabla_{a} \Sigma^{-1} + \frac{1}{2} \nabla^{a} \Sigma^{-1} \nabla_{a} \Sigma^{-1}=-\frac{1}{2} \nabla^{a} \Sigma^{-1} \nabla_{a} \Sigma^{-1}\,,
\ee
and therefore 
\be
-\frac{1}{4} \Sigma^{2}  \nabla^{a} \Sigma^{-1} \nabla_{a} \Sigma^{-1} - \frac{1}{2} \Sigma^{2} L^{a} \nabla_{a} \Sigma^{-1} \hateq0\,,
\ee
Hence, we get
\begin{align} \label{eq:final}
n^{b} l^{a} \nabla_{a} l_{b}&\hateq - \frac{1}{2} (\Sigma L^{a} \nabla_{a} (N^{b} \nabla_{b} \Sigma^{-1}) 
+ \Sigma^{2} \nabla_{a} \Sigma^{-1} \nabla^{a}\Sigma^{-1} - \Sigma \bar{L}^{b} N^{a} \nabla_{a} \nabla_{b} \Sigma^{-1})\notag \\
&\hateq -\frac{1}{2}(\Sigma L^{a} \nabla_{b} \Sigma^{-1} \nabla_{a} N^{b} +  \Sigma^{2} \nabla_{a} \Sigma^{-1} \nabla^{a}\Sigma^{-1}) \notag \\
&\hateq -\frac{1}{2} (\Sigma L^{a} \nabla_{b} \Sigma^{-1} (\frac{1}{2} n^{b} \nabla_{a}\Sigma + \Sigma \delta_{a}{}^{b} ) +  \Sigma^{2} \nabla_{a}\Sigma^{-1} \nabla^{a} \Sigma^{-1}) \notag\\
& \hateq -\frac{1}{2} L^{a} \nabla_{a} \Sigma - \frac{1}{2} \Sigma^{2}  L^{a} \nabla_{a} \Sigma^{-1} - \frac{1}{2}  \Sigma^{2} \nabla_{a} \Sigma^{-1} \nabla^{a} \Sigma^{-1}\notag\\
& \hateq -\frac{1}{4} \Sigma^{2} \nabla_{a} \Sigma^{-1} \nabla^{a} \Sigma^{-1} + \frac{1}{4} \Sigma^{2}  \nabla_{a} \Sigma^{-1} \nabla^{a} \Sigma^{-1} - \frac{1}{2}  \Sigma^{2} \nabla_{a} \Sigma^{-1} \nabla^{a} \Sigma^{-1} \,,
\end{align}
where in going to the second line, we used the fact that $L^{a} \hateq \bar{L}^{a}$ and in the subsequent steps we used \cref{cond:Sigma-choice,eq:N-defn,eq:l-der-sigmainv,eq:derninphi2}. We therefore see that                    
\be \label{eq:affin-phi2}
    n^{b} l^{a} \nabla_{a} l_{b} \hateq -\frac{1}{2} \Sigma^{2} \nabla_{a} \Sigma^{-1} \nabla^{a} \Sigma^{-1}\,.
\ee 


\bibliographystyle{JHEP}
\bibliography{BMS-matching}      
\end{document}